\documentclass[11pt]{article}

\usepackage{epsfig}
\usepackage{times}
\usepackage{epstopdf}
\usepackage{amsmath}
\usepackage{amssymb}
\usepackage[left=1.75cm,right=1.75cm,top=3.0cm,bottom=3.0cm]{geometry}
\usepackage{setspace}

\begin{document}

\bigskip\bigskip
\centerline {\Large \bf {Post-Newtonian Conservation Laws in Rigid Quasilocal Frames}}
\bigskip\bigskip
\centerline{\large Paul L. McGrath$^{a}$, Melanie Chanona$^{a}$, Richard J. Epp$^{a}$, Michael J. Koop$^b$ and Robert B. Mann$^{a,c}$}
\bigskip\bigskip
\centerline{${}^a$ \em Department of Physics and Astronomy, University of Waterloo, Waterloo, Ontario N2L 3G1, Canada}
\vspace{0.2 cm}
\centerline{${}^b$ \em Department of Physics, The Pennsylvania State University, PA 16802-6300, USA}
\vspace{0.2 cm}
\centerline{${}^c$ \em Perimeter Institute for Theoretical Physics, Waterloo, Ontario N2L 2Y5, Canada}
\vspace{0.3cm}
\centerline{\em  pmcgrath@uwaterloo.ca, mchanona@uwaterloo.ca, rjepp@uwaterloo.ca, mjk421@psu.edu, rbmann@uwaterloo.ca}
\bigskip\bigskip

\newcommand{\on}[2]{\stackrel{\phantom{}_{#1}}{#2}}

\begin{abstract}
In recent work we constructed completely general conservation laws for energy \cite{EMM2012} and linear and angular momentum \cite{EMM2013} of extended systems in general relativity based on the notion of a rigid quasilocal frame (RQF). We argued at a fundamental level that these RQF conservation laws are superior to conservation laws based on the local stress-energy-momentum tensor of matter because (1) they do not rely on spacetime symmetries and (2) they properly account for both matter {\it and} gravitational effects. Moreover, they provide simple, exact, operational expressions for fluxes of gravitational energy and linear and angular momentum. In this paper we derive the form of these laws in a general first post-Newtonian (1PN) approximation, and then apply these approximate laws to the problem of gravitational tidal interactions. We obtain formulas for tidal heating and tidal torque that agree with the literature, but without resorting to the use of pseudotensors. We describe the physical mechanism of these tidal interactions not in the traditional terms of a Newtonian gravitational force, but in terms of a much simpler and universal mechanism that is an exact, quasilocal manifestation of the equivalence principle in general relativity. As concrete examples, we look at the tidal heating of Jupiter's moon Io and angular momentum transfer in the Earth-Moon system that causes a gradual spin-down of the Earth and recession of the Moon. In both examples we find agreement with observation.
\end{abstract}

\section{Introduction and Summary} \label{secIntroductionPN}

In a dynamical curved spacetime the concepts of energy and momentum (linear and angular) of a spatially extended system are rather subtle for two main reasons. First, they are not local: there is no such thing as an energy or momentum per unit volume, which when integrated over a finite volume yields the total energy or momentum inside. As we argue below, this applies to both gravity {\it and} matter. Second, they are frame-dependent constructs. Newtonian space-time admits inertial reference frames, or more generally, arbitrarily accelerating or rotating frames in rigid motion, relative to which we can define the energy, momentum, and angular momentum of an extended system. The key concept is not inertial frame, but rigid frame, in which the distances between all nearest-neighbouring pairs of observers comprising the frame are constant in time. While special relativity admits some frames in rigid motion, e.g., those with arbitrary time-dependent acceleration but {\it no} rotation~\cite{SalzmanTaub1954}, general relativity generically admits none. Most relativists have thus retreated to notions of total energy, momentum and angular momentum for isolated systems in spacetimes that admit asymptotic symmetries at infinity, and rely on the pre-general relativistic practice of employing spacetime symmetries to construct conservation laws.

In previous work we have proposed breaking from this tradition. Our approach is based on the Brown and York quasilocal stress-energy-momentum tensor~\cite{BY1993} as a solution to the non-locality problem, together with our notion of a rigid quasilocal frame (RQF)~\cite{EMM2012,EMM2013,EMM2009,EMM2011} as a solution to the rigidity problem. We will briefly review these two in turn.

When a spacetime manifold, $\mathcal{M}$, has a non-dynamical (absolute) background metric, $g$, the total action for a system containing dynamical matter fields, $\varphi$, is just the matter action, $I_{\rm mat}[g,\varphi]$. Assuming for simplicity minimal coupling of matter to gravitation, the total stress-energy-momentum tensor of the system is found by computing the functional derivative of the total action with respect to the metric:
\begin{equation}\label{T_matter}
2\delta_g \, I_{\rm mat}[g,\varphi]=\int_\mathcal{M}\epsilon_\mathcal{M} \,T^{ab}\,\delta g_{ab},
\end{equation}
where $\epsilon_\mathcal{M}$ is the volume form on $\mathcal{M}$. This equation says that the total stress-energy-momentum tensor of the system is {\it local}, and is just the usual matter stress-energy-momentum tensor,  $T^{ab}$, which does not capture gravitational stress-energy-momentum. While $T^{ab}$ is covariantly conserved in general relativity, any conservation law constructed from $\nabla_a T^{ab}=0$ will be homogeneous in $T^{ab}$, and thus essentially blind to gravitational physics. In general relativity, on the other hand, the metric is regarded as a dynamical field, and so the total action must include also the action of the gravitational field. Using the usual first order gravitational action, and again for matter minimally coupled to gravity, we have:
\begin{equation}\label{T_matter_and_gravity}
2\delta_g \, I_{\rm mat+grav}[g,\varphi]=\int_\mathcal{M}\epsilon_\mathcal{M} \, \left(T^{ab}-\frac{1}{\kappa}G^{ab}\right)\,\delta g_{ab}+\int_{\cal B}\epsilon_{\cal B} \, \left(-\frac{1}{\kappa}\Pi^{ab}\right)\,\delta \gamma_{ab},
\end{equation}
where $\kappa=8\pi G/c^4$ and $G^{ab}$ is the Einstein tensor. This equation says that when we promote the metric to a dynamical field, the total {\it local} stress-energy-momentum tensor of the system is the sum, $T^{ab}-\frac{1}{\kappa}G^{ab}$, which is just zero by the Einstein equation. In other words, there is no non-trivial local notion of total stress-energy-momentum in general relativity, a statement that applies to both gravitational {\it and} matter fields. What saves us is the boundary term. In equation~(\ref{T_matter_and_gravity}), $\mathcal{B}$ is the boundary of $\mathcal{M}$, $\epsilon_\mathcal{B}$ and $\gamma_{ab}$ are the volume form and metric induced on $\mathcal{B}$, and $\Pi^{ab}$ is the gravitational momentum canonically conjugate to $\gamma_{ab}$.\footnote{In equation~(\ref{T_matter_and_gravity}), $\mathcal{B}$ is the timelike worldtube component of the boundary of $\mathcal{M}$. We have omitted the spacelike ``end caps" and "edge" components on the right-hand side of this equation, which do not play a role in our discussion.} Brown and York advocate replacing the local matter stress-energy-momentum tensor $T^{ab}$ with the quasilocal matter-plus-gravity stress-energy-momentum tensor $T_\mathcal{B}^{ab} = -\frac{1}{\kappa} \Pi^{ab}$~\cite{BY1993}.

While this addresses the non-locality problem, as we have argued especially in reference~\cite{EMM2013} we still need a solution to the rigidity problem in order to be able to construct sensible conservation laws for energy, momentum and angular momentum of extended systems. Just as the non-locality problem has a quasilocal solution, so does the rigidity problem. Rigid motion in a local sense requires that the orthogonal distance between all nearest-neighbour pairs of a three-parameter (volume-filling) family of observers comprising the frame remains constant in time. This represents six differential constraints on three degrees of freedom, and generically has no solutions. Rigid motion in a quasilocal sense restricts the rigidity condition to only the two-parameter family of observers on the boundary of the volume in question, which we assume is topologically a two-sphere. This represents only three differential constraints on the same three degrees of freedom, which we have argued especially in reference~\cite{EMM2011} can generically be solved. Moreover, the solution space (at least perturbatively) is in one-to-one correspondence with the six-parameter family of arbitrarily accelerating and rotating rigid frames we are familiar with in Newtonian mechanics. In such rigid quasilocal frames the four-velocity of the RQF observers satisfies a certain spatially-projected Killing vector condition that is invariant under an arbitrary functional rescaling of the four-velocity, and which is used in place of a spacetime timelike Killing vector to construct a sensible energy conservation law for extended systems. Moreover, a two-sphere always admits precisely six conformal Killing vector (CKV) fields that represent the action of the Lorentz group on the sphere, and which are used in place of spacetime spacelike Killing vectors to construct sensible linear momentum (boost CKV) and angular momentum (rotation CKV) conservation laws for extended systems. We emphasize that these conservation laws do not require any spacetime symmetries, and do not make use of a stress-energy-momentum tensor.

Geometrically, the RQF congruence has zero expansion and zero shear. Physically, the RQF observers are accelerating in such a way that the locally-measured size and shape of the system boundary do not change, despite the flow of gravitational or matter fluxes through the boundary. The RQF observers can thus consider themselves to be ``at rest" in an (inhomogeneous and time-dependent) gravitational field, in the sense of the equivalence principle. This results in a clean identification of energy-momentum fluxes since fluxes due merely to changes in the size or shape of the boundary are eliminated. The result is simple, exact, operational expressions for fluxes of gravitational energy, linear momentum and angular momentum that are direct quasilocal analogues of the corresponding ``bulk" terms that appear in a local conservation law to account for non-inertial motion of the frame. In other words, these gravitational fluxes are precisely the difference between rigid frames (that can be arbitrarily accelerating or rotating) and inertial frames, and as we shall see, represent an exact, quasilocal manifestation of the equivalence principle in general relativity.

The main purpose of this paper is to provide general expressions for these gravitational fluxes and associated RQF conservation laws in a first post-Newtonian (1PN) approximation, and then apply these approximate results in the context of tidal gravitational interactions. Our results for tidal power and torque agree with generally accepted results based on the use of stress-energy-momentum pseudotensors, and provide a long-sought geometrical justification of the latter. Moreover, establishing this connection with accepted results in the slow-motion weak-field approximation allows us to argue that the RQF approach really does provide an exact, non-perturbative explanation of the physical mechanism behind important gravitational effects such as tidal power and torque---a mechanism based not on the Newtonian gravitational force, but on the equivalence principle and geometry.

We begin in~{\S}\ref{secConservationLaws} with a brief review of the notion of an RQF, and how the RQF concept gives rise to a set of completely general conservation laws for energy, linear momentum, and angular momentum of extended systems. In~{\S}\ref{secPNRQFLaws} we analyze these conservation laws in the post-Newtonian context. This is done by starting with a general first post-Newtonian metric and embedding in it an RQF to derive a new metric describing the post-Newtonian spacetime as seen by RQF observers. It is in this metric that we evaluate the RQF conservation laws of the previous section and ultimately derive expressions for the time rate of change of the energy and angular momentum inside the RQF in terms of gravitational fluxes through the RQF boundary.\footnote{\label{1PN}Technically, we are not using the full first post-Newtonian approximation. We are dropping the 1PN terms in the space-space components of the metric. This tremendously simplifies the analysis, and is sufficient to obtain useful results in the case of energy and angular momentum conservation. The case of linear momentum conservation requires the 1PN terms in the space-space components of the metric to determine the gravitational flux of linear momentum. This is a more difficult analysis, the results of which will be presented elsewhere.}

In {\S}\ref{secTidalExamples} we apply these results to tidal gravitational interactions. Specifically, we consider the spacetime of a body represented by a general multipole expansion immersed in a weak external gravitational field and derive formulas that allow us to analyze the work done by tidal interactions as well as the transfer of angular momentum due to tidal torques. We then use these equations to analyze tidal interactions in the familiar arena of our solar system.  In particular, by putting an RQF around Jupiter's moon Io, we compute the energy transferred to Io due to the interaction between Io's quadrupole moment and the tidal forces of Jupiter. Additionally, we calculate the tidal torque that the Moon exerts on the Earth. This torque mines the Earth of angular momentum and causes the Moon to recede in its orbit. In both cases, we compute numerical values for these effects and in both cases we find solid agreement with observation. We emphasize again that while some of these results have been previously obtained using other approaches, the RQF approach does not rely on pseudotensors or spacetime symmetries: it is a manifestly geometrical approach that provides a qualitatively different (and simpler) explanation of gravitational interactions. {\S}\ref{secSummary} contains a summary and conclusions.

\section{Conservation Laws in Rigid Quasilocal Frames}\label{secConservationLaws}

\subsection{Definition of a Rigid Quasilocal Frame}

We briefly review the definition of a rigid quasilocal frame and several of its features that will be important in this work. Consider a smooth four-dimensional manifold with a Lorentzian metric, $g_{ab}$,  and an associated covariant derivative operator $\nabla_a$ and volume form $\epsilon_{abcd}$. In this spacetime, let $\mathcal B$ denote a two-parameter congruence of timelike worldlines with topology $\mathbb{R}\times \mathbb{S}^{2}$, i.e., a timelike worldtube representing the history of a two-sphere's-worth of observers bounding a finite spatial volume. Let $u^a$ be the future-directed unit vector field tangent to this congruence, representing the observers' four-velocity. The metric $g_{ab}$ induces on $\mathcal B$ a spacelike outward-directed unit normal vector field, $n^{a}$, and a Lorentzian three-metric, $\gamma_{ab}=g_{ab}-n_{a}n_{b}$, with associated covariant derivative operator $D_a$ and volume form $\epsilon_{abc}=n^d \epsilon_{abcd}$. Let $\sigma_{ab}=\gamma_{ab}+\frac{1}{c^2} u_a u_b$ denote the RQF observers' spatial two-metric (measuring orthogonal distance between the worldlines of the congruence) with associated covariant derivative operator $\hat{D}_a$ and volume form $\epsilon_{ab}=\frac{1}{c}u^c n^d \epsilon_{abcd}$. The time development of the congruence is then described by the tensor field $\theta_{ab}= \sigma_a^{\phantom{a}c}\sigma_b^{\phantom{b}d}\nabla_c u_d$. We decompose $\theta_{ab}$ into the local {\it expansion}, $\theta = \sigma^{ab}\theta_{ab}$, {\it shear}, $\theta_{< ab >} = \theta_{(ab)}-\frac{1}{2}\theta\sigma_{ab}$, and {\it twist}, $\nu=\frac{1}{2}\epsilon^{ab}\theta_{ab}$, of the worldlines.

A {\it rigid quasilocal frame} (RQF) is defined as a congruence of the type described above but with zero expansion and zero shear. In other words, the local size and shape of the boundary of the finite spatial volume---as measured by the RQF observers using the orthogonal (``radar-ranging") distance between nearest neighbours---do not change with time:
\begin{equation} \label{RigidityConditionPN}
\theta = 0\;\;\;{\rm and}\;\;\;\theta_{<ab>}=0\;\;\;\Longleftrightarrow\;\;\;\theta_{(ab)}=0 .
\end{equation}
These conditions amount to three differential constraints that rigidly fix the intrinsic spatial two-geometry perceived by the two-sphere's-worth of RQF observers. For future reference, we also define the observers' four-acceleration $a^a=u^b\nabla_b u^a$, whose projection tangential to $\mathcal B$ is $\alpha^a=\sigma^a_{\phantom{a}b}a^b$ (an intrinsic geometrical variable), and projection normal to $\mathcal B$ is $n\cdot a = n_a a^a$ (an extrinsic geometrical variable).

To further clarify the RQF construction, and to establish notation for some of the results in subsequent sections, let us introduce a coordinate system adapted to the congruence. Thus, let two functions $x^i$ on ${\mathcal B}$ locally label the observers, i.e., the worldlines of the congruence. Let the function $t$ on ${\mathcal B}$ denote a time parameter such that the surfaces of constant $t$ form a foliation of ${\mathcal B}$ by two-surfaces with topology $\mathbb{S}^{2}$. Collect these three functions together as a coordinate system, $x^\mu = (t,x^i)$, and set $u^\mu = \delta_t^\mu /N$, where $N$ is a lapse function ensuring that $u\cdot u=-c^2$. The general form of the induced metric $\gamma_{ab}$ then has adapted coordinate components:

\begin{equation} \label{eq:InducedMetricPN}
\gamma_{\mu\nu} = \left(
\begin{array}{cc}
-c^2 N^2 & N u_j \\
N u_i & \sigma_{ij}-\frac{1}{c^2}u_i u_j
\end{array}
\right).
\end{equation}
Here $\sigma_{ij}$ and the shift covector $u_i$ are the $x^i$ coordinate components of $\sigma_{ab}$ and $u_{a}$, respectively. Note that $\sigma_{ij}\,dx^i\,dx^j$ is the radar-ranging, or orthogonal distance between infinitesimally separated pairs of observers' worldlines, and it is a simple exercise to show that the RQF rigidity conditions in equation~(\ref{RigidityConditionPN}) are equivalent to the three conditions $\partial\sigma_{ij}/\partial t =0$. In other words, an RQF is a {\it rigid} frame in the sense that each observer sees himself to be permanently at rest with respect to his nearest neighbours. The local proper accelerations and rotations that observers must undergo to maintain this rigidity then encode information about the spacetime the RQF is immersed in. We discuss the generic existence of RQFs in earlier papers, especially reference~\cite{EMM2011}.

Assuming the RQF conditions are satisfied, we are free to perform a time-{\it in}dependent coordinate transformation amongst the $x^i$ (a relabelling of the observers) such that $\sigma_{ij}$ takes the form $\sigma_{ij}=\Omega^2 \, \mathbb{S}_{ij}$, where $\Omega^2$ is a time-independent conformal factor encoding the size and shape of the rigid two-boundary, and $\mathbb{S}_{ij}$ is the standard metric on the unit round sphere. For our purposes, we will take the observers' two-geometry to be that of a round sphere with areal radius $r$, with the observers labelled by the standard spherical coordinates $x^i=(\theta,\phi)$, so that $\mathbb{S}_{ij}=\text{diagonal}(1,\sin^2\theta)$ and $\Omega=r$.

The intrinsic geometrical degrees of freedom of the RQF are encoded in the observers' coordinate-independent proper acceleration tangential to $\mathcal B$ ($\alpha^{a}$ defined above). In addition to $\alpha^a$ we are free to specify the twist, $\nu$, on one cross section of $\mathcal B$. In the adapted coordinate system, these geometrical degrees of freedom are related to $N$ and $u_i$ through
\begin{equation}\label{eq:ObserversAccelerationPN}
\alpha_{i}=\frac{1}{N}\,\dot{u}_{i}+c^2\partial_{i}\ln N,
\end{equation}
and
\begin{equation}\label{eq:nuPN}
\nu = \frac{1}{2}\epsilon^{ij}(\partial_i u_j - \frac{1}{c^2}\alpha_i u_j ).
\end{equation}
In these equations, an over-dot denotes partial derivative with respect to $t$, and $\partial_{i}$ denotes partial derivative with respect to $x^{i}$. Also, $\epsilon^{ij}$ are the $x^i$ adapted coordinate components of $\epsilon^{ab}$. Under a time parameter re-foliation of ${\mathcal B}$, $N$ and $u_i$ change in a complementary way such that $\alpha_{a}$ and $\nu$ transform covariantly. Sometimes it is convenient to choose the foliation of ${\mathcal B}$ such that the exact (gradient) part of $u_i$ vanishes, i.e., the RQF observers are as close to being ``at rest" (their four-velocity is as close to being orthogonal to surfaces of constant $t$) as the geometry allows.

\subsection{Conservation Laws in Rigid Quasilocal Frames}

As mentioned in the {\S}\ref{secIntroductionPN}, the Brown and York~\cite{BY1993} quasilocal stress-energy-momentum tensor, defined on the boundary, $\mathcal{B}$, is given by $T_\mathcal{B}^{ab} = -\frac{1}{\kappa} \Pi^{ab}$, where $\Pi^{ab}=K^{ab}-K\gamma^{ab}$ is the gravitational momentum canonically conjugate to $\gamma_{ab}$, $K_{ab}=\gamma_{a}^{\phantom{a}c}\nabla_c n_b$ is the extrinsic curvature of $\mathcal{B}$, and $\kappa=8\pi G/c^4$. Using $D_a$, the covariant derivative operator induced on $\mathcal{B}$, we start with the following identity:
\begin{equation}\label{differential_quasilocal_conservation_lawPN}
D_a (T_\mathcal{B}^{ab}\psi_b )=(D_a T_\mathcal{B}^{ab})\psi_b + T_\mathcal{B}^{ab}D_{(a}\psi_{b)},
\end{equation}
where $\psi^a$ is an arbitrary vector field tangent to $\mathcal{B}$. Appropriate choices of $\psi^a$ (made below) will lead to conservation laws for either energy, linear momentum, or angular momentum. Integrating equation~(\ref{differential_quasilocal_conservation_lawPN}) over $\Delta\mathcal{B}$, a portion of $\mathcal{B}$ bounded by initial-time and final-time two-surfaces $\mathcal{S}_i$ and $\mathcal{S}_f$, we have:
\begin{equation} \label{integrated_quasilocal_conservation_lawPN}
\frac{1}{c} \int\limits_{\mathcal{S}_f - \mathcal{S}_i}  d{\mathcal{S}}\,  T_{\mathcal B}^{ab} u^{\mathcal{S}}_a \psi_b = \int\limits_{\Delta\mathcal{B}}  d \mathcal{B} \, \left[ T^{ab} n_a \psi_b - T_{\mathcal B}^{ab} D_{(a} \psi_{b)} \right].
\end{equation}
Here we used the Gauss-Codazzi identity, $D_a \Pi^{ab}=n_a G^{ac} \gamma_c^{\phantom{c}b}$, where $G^{ab}$ is the Einstein tensor, and the Einstein equation, $G^{ab}=\kappa T^{ab}$, which is how the local matter stress-energy-momentum tensor enters into what is otherwise a geometrical identity. Also, $\frac{1}{c}u_{\mathcal{S}}^a$ denotes the timelike future-directed unit vector field tangent to $\mathcal{B}$ and orthogonal to $\mathcal{S}_i$ and $\mathcal{S}_f$. Finally, we follow Brown and York again to resolve $T_{\mathcal B}^{ab}$ into components adapted to the $u^a$-observers:
\begin{equation} \label{BYDecompositionPN}
T_\mathcal{B}^{ab} = \frac{1}{c^2}u^a u^b\mathcal{E}+2u^{(a}\mathcal{P}^{b)}-\mathcal{S}^{ab}.
\end{equation}
Here $\mathcal{E}$, $\mathcal{P}^a$, and $\mathcal{S}^{ab}$ are the quasilocal energy, momentum, and stress surface densities, respectively (e.g., $\mathcal{E}$ has dimensions of energy per unit area).

Using this decomposition of $T_\mathcal{B}^{ab}$, and taking $\psi^a = u^a$ in equation (\ref{integrated_quasilocal_conservation_lawPN}), we arrive at the following completely general RQF energy conservation law \cite{EMM2012,EMM2011}:
\begin{align}\label{EnergyGCLPN}
\int\limits_{\mathcal{S}_f - \mathcal{S}_i}  d\hat{\mathcal{S}} \, & \left[ \mathcal{E} - \mathcal{P}^a v_a \right]  = - \int\limits_{\Delta{\mathcal B}}  N \, dt \, d\hat{\mathcal{S}}\, \left[  - T^{ab} n_a u_a + \alpha_a \mathcal{P}^a  \right].
\end{align}
Regarding the integration measures, we have extended the definition of $u_{\mathcal{S}}^a$ from $\mathcal{S}_i$ and $\mathcal{S}_f$ to all surfaces $\mathcal{S}_t$ of constant $t$ foliating $\Delta{\mathcal B}$ via the decomposition $u^a_\mathcal{S}=\gamma (u^a+ v^a)$, where $v^a$ is tangent to $\mathcal{B}$ and orthogonal to $u^a$, and $\gamma=(1-v^2/c^2)^{-1/2}$ is a Lorentz factor. Physically, $v^a$ represents the spatial two-velocity of fiducial observers who are ``at rest" with respect to $\mathcal{S}_t$ (their four-velocity $u^a_\mathcal{S}$ is orthogonal to $\mathcal{S}_t$) as measured by the RQF observers (whose four-velocity is $u^a$). Note that since $u^a_\mathcal{S}$ is orthogonal to $\mathcal{S}_t$, $v_i=-u_i$ in the adapted coordinate system introduced earlier. Also note that $\gamma d\mathcal{S} = d\hat{\mathcal{S}}$ is the surface element seen by the RQF observers, which is constant in time. In the case of a round sphere of areal radius $r$, $d\hat{\mathcal{S}} = r^2 \sin \theta \, d\theta \, d \phi$ in the adapted coordinate system.

Taking instead $\psi^a = -\frac{1}{c}\phi^a$ in equation (\ref{integrated_quasilocal_conservation_lawPN}), where $\phi^a$ can be either one of the boost or rotation CKVs mentioned in {\S}\ref{secIntroductionPN}, yields the following completely general RQF linear and angular momentum conservation law \cite{EMM2013}:
\begin{align}\label{MomentumGCLPN}
\int\limits_{\mathcal{S}_f - \mathcal{S}_i}  d\hat{\mathcal{S}} \, \left[   \left( \mathcal{P}^a + \frac{1}{c^2}\mathcal{S}^{ab} v_b\right) \phi_a \right] = - & \int\limits_{\Delta{\mathcal B}}  N \, dt \, d \hat{\mathcal{S}}\, \left[ \left( T^{ab} n_a \phi_b  + {\rm P} \hat{D}_{a} \phi^{a}\right) + \left(\frac{1}{c^2} \mathcal{E} \alpha^a - 2 \nu \epsilon^{ab} \mathcal{P}_b  \right) \phi_a \right].
\end{align}
The vectors $\phi^a$ satisfy the following properties: (1) they are spatial ($\phi^a=\sigma^a_{\phantom{a}b}\phi^b$); (2) they are ``stationary" ($\sigma^a_{\phantom{a}b}\mathcal{L}_u \phi^b=0$), i.e., they appear to the RQF observers to be constant vectors lying along their local space axes; and (3) they are CKVs ($\hat{D}_{(a}\phi_{b)}=\frac{1}{2}\sigma_{ab}\hat{D}_c \phi^c$). The quasilocal pressure, defined as ${\rm P}=\frac{1}{2}\sigma_{ab}\mathcal{S}^{ab}$, is a kind of gravitational ``force" per unit length between the RQF observers that plays the role (at least at lowest order in an areal radius expansion of a small RQF) of adding the ``missing" normal-normal component of matter stress to the normal-tangential matter stress term $T^{ab} n_a \phi_b$. In the case of a boost CKV (linear momentum), $\hat{D}_{a} \phi^{a}$ is an $\ell=1$ spherical harmonic which extracts the corresponding $\ell=1$ component of ${\rm P}$, i.e., a gradient of pressure resulting in an effective linear force. In the case of a rotation CKV (angular momentum), $\hat{D}_{a} \phi^{a}=0$---the normal-tangential matter stress term $T^{ab} n_a \phi_b$ is sufficient to represent the full matter torque. \cite{EMM2013}

The RQF energy and momentum conservation laws in equations~(\ref{EnergyGCLPN}) and (\ref{MomentumGCLPN}) are discussed in detail in references~\cite{EMM2012,EMM2013,EMM2011}. The important features for the present paper can be summarized as follows:
\begin{itemize}
\item Standard conservation laws, e.g., equations (2.2) and (2.3) in reference \cite{HT1985}, provide only {\it approximate} notions of mass, linear momentum, and angular momentum, and their conservation laws, valid in a certain {\it approximate} asymptotically flat region of the compact body in question (the ``buffer zone"). Moreover, they are based on pseudotensors, so there is always the worry about coordinate dependence of the analysis. In contrast, equations~(\ref{EnergyGCLPN}) and (\ref{MomentumGCLPN}) are {\it exact}, valid in any region in which an RQF exists, including strong-field regions. They are also purely geometrical, manifestly coordinate-independent equations. And they do not rely on any spacetime symmetries. Of course the drawback is that, in most practical cases, RQFs can be constructed only {\it approximately}, e.g., in the same sort of ``buffer zone" just mentioned. But at least {\it conceptually} we can make exact statements pertaining to various physical mechanisms at work in general relativity; in this paper, these statements are to do with transfer of energy and angular momentum in gravitational tidal interactions.
\item On the left-hand side of equations~(\ref{EnergyGCLPN}) and (\ref{MomentumGCLPN}) we see that $\mathcal{E}$ and $\mathcal{P}^a$ are augmented by certain stress terms associated with a non-zero $v^a$. As discussed in reference~\cite{EMM2013}, these are required to adjust for the case when the RQF observers are not ``at rest" with respect to the surfaces $\mathcal{S}_i$ and $\mathcal{S}_f$, i.e., their four-velocity is not orthogonal to $\mathcal{S}_i$ and $\mathcal{S}_f$. This is the general relativistic analogue of the solution to the famous ``$4/3$ problem" in the special relativistic theory of the electron. As we shall see, the stress term in equation~(\ref{MomentumGCLPN}) plays a non-trivial role in the definition of angular momentum in tidal interactions.
\item On the right-hand side of equations~(\ref{EnergyGCLPN}) and (\ref{MomentumGCLPN}) we see the usual matter energy, linear momentum, and angular momentum fluxes (augmented by the ${\rm P} \hat{D}_{a} \phi^{a}$ term in the case of linear momentum, discussed above). The remaining terms are precisely the ``bulk" terms one encounters in local conservation laws that account for non-inertial effects due to acceleration and rotation of the rigid frame, promoted to {\it exact} fluxes of gravitational energy, linear momentum, and angular momentum. These fluxes represent the difference between {\it rigid} frames and {\it inertial} frames of reference. As we shall see, they account entirely for fluxes of energy and angular momentum in gravitational tidal interactions.
\end{itemize}
	
\section{Post-Newtonian Expansion of the RQF Conservation Laws}\label{secPNRQFLaws}

We will now analyze the general RQF conservation laws above in the first post-Newtonian (1PN) approximation. To do this, we will need to find the metric for an RQF embedded in the standard post-Newtonian spacetime.  Recall that, in the post-Newtonian scheme, we deal with non-relativistic systems that are bound by weak mutual gravitational attraction amongst constituent particles so that kinetic energies are comparable to gravitational potential energies. This allows us to expand metric quantities in terms of a dimensionless parameter $\epsilon \sim V/c \sim \sqrt{GM/c^2R}$ where $V$, $M$, and $R$ are typical velocities, masses, and separation distances respectively of the particles comprising the system under study.  This expansion leads to the post-Newtonian metric, which can be found in many standard textbooks on general relativity (see \cite{Weinberg} for example).  In pseudo-Cartesian coordinates $X^A = (X^0=cT, X^I)$, $I=1,2,3$, this metric is given by
\begin{align}\label{WeinbergMetric}
g_{00} &= -1 - \frac{2\Phi}{c^2}  - \left(\frac{2\Phi^2}{c^4} + \frac{2\Psi}{c^4} \right) + \mathcal{O}(\epsilon^6), \nonumber \\
g_{0J} &= \frac{\zeta_J}{c^3} + \mathcal{O}(\epsilon^5), \nonumber \\
g_{IJ} &= \delta_{IJ} \left(1 - \frac{2 \Phi}{c^2}\right) + \mathcal{O}(\epsilon^4)
\end{align}
where $\Phi/c^2 \sim \mathcal{O}(\epsilon^2)$, $\zeta_J/c^3 \sim \mathcal{O}(\epsilon^3)$, and $\Psi/c^4 \sim \mathcal{O}(\epsilon^4)$ can be functions of all of the coordinates $X^A$. Technically, this is not quite a 1PN approximation---see footnote~\ref{1PN} on page~\pageref{1PN}, but is sufficient for our purposes.

To move to the RQF frame, with adapted coordinates $x^\alpha = (t,r,x^i = (\theta, \phi))$, we apply the transformation
\begin{align} \label{PNTransformation}
cT &= ct + \on{1}{f^0} +  \on{3}{f^0} +   \mathcal{O}(\epsilon^5), \nonumber \\
X^I &= r r^I + \on{2}{f^I} +   \mathcal{O}(\epsilon^4),
\end{align}
where $r^I = (\sin \theta\, \cos \phi  , \,\sin \theta \, \sin \phi , \,\cos \theta)$ are the usual direction cosines. The number $n$ above the functions $\on{n}{f^A}$ denotes the order in $\epsilon$ of that function.  Note that the transformation of the time coordinate involves only odd powers of $\epsilon$ because it must change sign under time-reversal, whereas the spatial transformation has only even powers to keep the same sign under time-reversal. The functions $\on{n}{f^0}$ allow for an arbitrary infinitesimal perturbation in the time foliation, while the three sets of functions $\on{n}{f^I}$ introduce enough freedom in spatial perturbations of the coordinate embedding to satisfy the three RQF conditions.

Before we actually solve the RQF equations and give the full metric resulting from the transformation above, let us make an observation that will simplify the end result.  Following the transformation (\ref{PNTransformation}), one finds $\on{1}g_{tr} = -c \partial_{r}\on{1}{f^0}$ and $\on{1}g_{tj} = - c \partial_{j}\on{1}{f^0}$.  However, for a general post-Newtonian RQF, i.e., equation (\ref{WeinbergMetric}), these metric components vanish at this order.  This is a result of the fact that, at zeroth order in $\epsilon$, the standard post-Newtonian spacetime has zero acceleration and rotation.
Thus we must take
\begin{align}
\partial_r \on{1}{f^0} = 0, \quad \partial_j \on{1}{f^0} = 0.
\end{align}
In other words, $\on{1}{f^0}$ can only have time dependence. With this simplification, the three RQF conditions in equation~(\ref{RigidityConditionPN}) can be shown to reduce to
\begin{align}\label{PNRQFEquation}
0 = 2r^2 \left( \frac{1}{r}\mathbb{B}_{I(i}\mathbb{D}_{j)}\on{2}{f^I} - \frac{\Phi}{c^2} \mathbb{S}_{ij} \right) + \mathcal{O}(\epsilon^4),
\end{align}
where we have introduced the covariant derivative operator, $\mathbb{D}_i$, associated with the unit two-sphere metric $\mathbb{S}_{ij}$, as well as the three boost CKVs $\mathbb{B}^I_i = \mathbb{D}_i r^I$.  It is straightforward to see that a particular solution to these three differential equations is given by $\on{2}{f^I_{\rm p}} = r r^I \Phi/c^2$.  We are also free to add to it the general homogeneous solution, $\on{2}{f^I_{\rm h}} = \alpha^I (t,r) + \epsilon^{I}_{\phantom{I}JK} r^J \beta^K (t,r)$, where $\epsilon_{IJK}$ is the alternating symbol.  Here, $\alpha^I (t,r)$ and $\beta^I (t,r)$ are six arbitrary functions of time (for a given $r$) that impart arbitrary $\ell=1$ acceleration and rotation to the RQF. They are the quasilocal analogues of the six degrees of freedom of arbitrarily accelerating and rotating rigid frames we are familiar with in Newtonian mechanics. However, in keeping with the fact that, at zeroth order in $\epsilon$, the standard post-Newtonian spacetime is inertial, we will suppress this freedom and take
\begin{equation}\label{RQFSolution}
\on{2}{f^I} = r r^I \frac{\Phi}{c^2},
\end{equation}
but will return to this point in {\S}\ref{secTidalExamples}. In this equation, $\Phi$ is a function of the RQF coordinates through the zeroth order version of equation~(\ref{PNTransformation}), i.e., $T=t$ and $X^I=rr^I$.

It is now straightforward to show that the metric for an RQF embedded in the post-Newtonian spacetime given in equation~(\ref{WeinbergMetric}) is
\begin{align} \label{RQFWeinberg}
g_{tt} &= -c^2 -2 \bigg[ \Phi + c \dot{\on{1}{f^0}} \bigg] - 2 \bigg[ \frac{\Psi}{c^2} + \frac{\Phi^2}{c^2} + \frac{r}{2 c^2} (\Phi^2)^\prime + \frac{2}{c} \Phi (\dot{\on{1}{f^0}}) + \frac{1}{c} \on{1}{f^0} \dot{\Phi}  + \frac{1}{2}(\dot{\on{1}{f^0}})^2 + c(\dot{\on{3}{f^0}}) \bigg]  + \mathcal{O}(\epsilon^6), \nonumber \\
g_{tr} &= \bigg[ \frac{\zeta}{c^2} + \frac{r}{c^2} \dot{\Phi} - c \on{3}{f^{0}}^\prime \bigg] + \mathcal{O}(\epsilon^5), \nonumber \\
g_{tj} &= \bigg[ \frac{r}{c^2} \zeta_j  - c \mathbb{D}_j \on{3}{f^0} \bigg] + \mathcal{O}(\epsilon^5), \nonumber \\
g_{rr} &= 1 + \big[ \frac{2 r}{c^2} \Phi^\prime \big] + \mathcal{O}(\epsilon^4), \nonumber \\
g_{rj} &= \big[ \frac{r}{c^2} \mathbb{D}_j \Phi  \big] + \mathcal{O}(\epsilon^4), \nonumber \\
g_{ij} &= r^2 \mathbb{S}_{ij} + \mathcal{O}(\epsilon^4),
\end{align}
where $\Phi$, $\zeta_J$, and $\Psi$ are now functions of the RQF coordinates $x^\alpha$ through the zeroth order version of equation~(\ref{PNTransformation}), and we have adopted a simplified notation of denoting radial derivatives with a prime, $\partial_r f  = f^\prime$, and time derivatives with a dot, $\partial_t f = \dot{f}$.  It is also important here to remember that time derivatives carry an order in $\epsilon$ since $\frac{1}{c}\frac{\partial}{\partial t} \sim \frac{v}{c} \frac{\partial}{\partial x^I} \sim \epsilon$.  For convenience, we have also decomposed $\zeta_J$ into a radial part, $\zeta = r^I \zeta_I$, and a part tangential to the RQF two-sphere, $\zeta_i := \mathbb{B}^I_i \zeta_I$.

It will be useful here to collect a few results that will recur throughout our analysis of the conservation laws. First, from the metric above, it is straightforward to write down the shift covector for the RQF observers,
\begin{align}\label{VelocityPN}
u_j = \frac{1}{N} g_{tj} = c \left[  r \frac{ \zeta_j}{c^3}  - \mathbb{D}_j \on{3}{f^0} \right] + \mathcal{O}(\epsilon^5).
\end{align}
Using the shift covector and equation (\ref{eq:ObserversAccelerationPN}) we can compute the tangential acceleration that the RQF observers must undergo to maintain rigidity,
\begin{align}\label{AccelPN}
\alpha_j = \bigg[ \mathbb{D}_j \Phi \bigg] + \bigg[ \frac{r}{2c^2} \mathbb{D}_j (\Phi^2)^\prime + \frac{r}{c^2} \dot{\zeta}_j + \frac{1}{c^2} \mathbb{D}_j \Psi + \frac{1}{c} \on{1}{f^0} \mathbb{D}_j \dot{\Phi} ) \bigg] + \mathcal{O}(\epsilon^6),
\end{align}
A lengthy calculation gives the quasilocal momentum density of the RQF,
\begin{align}\label{QuasilocalMomentum}
\mathcal{P}_j = \frac{1}{c^4 \kappa} \bigg[ \frac{1}{2} (r\zeta_j)^\prime + \frac{1}{4} \mathbb{D}_j \left( r \zeta^\prime + \mathbb{D}^k \zeta_k \right) \bigg] + \mathcal{O}(\epsilon^3).
\end{align}
Note that $c^4 \kappa \sim G$, so it is easy to see that $|\mathcal{P}|\times {\rm Area} \sim MV\epsilon$ at lowest order, where $MV$ represents a typical momentum in the system.

In previous work \cite{EMM2011}, we found that the equivalence principle can be used to relate the acceleration and quasilocal momentum density to effective gravitoelectricmagnetic (GEM) fields.  Let us make use of that idea now; the reasons for doing this will become apparent later on when we look at the flux of gravitational energy.  Thus we define the GEM potentials
\begin{align}
\phi^{\mathtt{GEM}} &= \Phi +\left[\frac{r}{2c^2} (\Phi^2)^\prime +\frac{1}{c^2} \Psi + \frac{1}{c} \on{1}{f^0} \dot{\Phi} \right]  + \mathcal{O}(\epsilon^6), \nonumber \\
A_I^{\mathtt{GEM}} &= - \frac{1}{4c} (r \zeta^\prime + \mathbb{D}^k \zeta_k ) r_I + \frac{1}{2c} \zeta_i \mathbb{B}_I^{i} + \mathcal{O}(\epsilon^5).
\end{align}
The acceleration in equation~(\ref{AccelPN}) can then be identified with the gravitoelectric field projected tangentially to the RQF sphere by the relation
\begin{align}\label{GravitoelectricPN}
e^{\mathtt{GEM}}_j = \mathbb{B}^J_j E^{\mathtt{GEM}}_J  = - \mathbb{D}_j \phi^{\mathtt{GEM}} - \mathbb{B}^J_j \dot{A}_J^{\mathtt{GEM}}  = - \alpha_j .
\end{align}
Similarly, the quasilocal momentum density can be identified with the tangential part of the gravitomagnetic field via
\begin{align}\label{GravitomagneticPN}
b^{\mathtt{GEM}}_j = \mathbb{B}^J_j B^{\mathtt{GEM}}_J  = \mathbb{B}^J_j \epsilon_J^{\phantom{J}KL} \partial_K A_L^{\mathtt{GEM}}  = - \frac{c^3 \kappa}{r} \mathbb{E}_j^{\phantom{j}k} \mathcal{P}_k ,
\end{align}
where $\partial_I$ denotes differentiation with respect to $x^I = r r^I$, and $\mathbb{E}_{ij}$ is the volume element associated with the metric $\mathbb{S}_{ij}$.  Lastly, it is worth noting that the twist of the congruence of RQF observers, see equation~(\ref{eq:nuPN}), is related to the radial part of the gravitomagnetic field:
\begin{align}\label{TwistPN}
r^J B^{\mathtt{GEM}}_J = \frac{1}{2 c r} \mathbb{E}^{ij} \mathbb{D}_i \zeta_j + \mathcal{O}(\epsilon^5)=c\nu .
\end{align}
The twist starts at order $\epsilon^3$ because we have followed the usual post-Newtonian approach and assumed that, at zeroth order in $\epsilon$, the spacetime is inertial. Thus, it is only non-zero once the effects of rotational frame-dragging arise.

Finally, one can show that the quasilocal energy density is given by
\begin{align}\label{QuasilocalEnergy}
\mathcal{E} = -\frac{2}{\kappa r} + \frac{1}{c^2 \kappa} \bigg[  2 \Phi^\prime +\frac{1}{r} \mathbb{D}^2 \Phi \bigg] + \mathcal{O}(\epsilon^2),
\end{align}
where $\mathbb{D}^2 = \mathbb{S}^{ij} \mathbb{D}_i  \mathbb{D}_j$. The energy density, in general, will involve vacuum, matter, and gravitational contributions. The first term on the right-hand side of
equation~(\ref{QuasilocalEnergy}) is the quasilocal vacuum energy density, which we will denote as
\begin{align}\label{EVacuum}
\mathcal{E}_{\mathtt vac} = - \frac{2}{\kappa r}.
\end{align}
Note that the vacuum term is essentially of order $1/\epsilon^{2}$. The remaining two terms are matter contributions beginning at zeroth order in $\epsilon$ and, for later reference, we define
\begin{align}\label{EMatter}
\mathcal{E}_{\mathtt mat} = \frac{1}{c^2 \kappa} \bigg[  2 \Phi^\prime +\frac{1}{r} \mathbb{D}^2 \Phi \bigg] + \mathcal{O}(\epsilon^2).
\end{align}
Notice that, for the simplest gravitational potential $\Phi = - G M / r$, this gives $\mathcal{E}_{\mathtt mat} = M c^2 / 4 \pi r^2$. Integrated over the RQF sphere, this quasilocal energy density gives a total matter energy equal to $Mc^2$, just as one would expect.  Note that, at these low orders, contributions to the quasilocal energy density due to gravitational effects (outside of $\mathcal{E}_{\mathtt vac}$) do not show up.

\subsection{Post-Newtonian Conservation Laws}

Working with the metric given in equation~(\ref{RQFWeinberg}) above, we can now evaluate equations~(\ref{EnergyGCLPN}) and (\ref{MomentumGCLPN}) to obtain conservation laws for energy, linear momentum, and angular momentum in the lowest order post-Newtonian limit.

\subsubsection*{Energy}

The integral on the left-hand side of equation~(\ref{EnergyGCLPN}) gives the change in the energy inside the RQF between the surfaces of simultaneity $\mathcal{S}_i$ and $\mathcal{S}_f$, including the term $-\mathcal{P}^a v_a$ required to adjust for the motion of the RQF observers relative to $\mathcal{S}_i$ and $\mathcal{S}_f$. However, inspection of equations~(\ref{VelocityPN}) and (\ref{QuasilocalMomentum}), and the fact that $v_i=-u_i$, reveals that $\mathcal{P}^a v_a \sim \mathcal{O}(\epsilon^4)$. On the other hand, $\mathcal{E}$ is only known at vacuum and zeroth order in $\epsilon$---see  equation~(\ref{QuasilocalEnergy}). Thus, the left-hand side of the energy conservation law involves only the lowest order matter contribution:
\begin{align}
\Delta\mathrm{E}_{\mathtt{RQF}} = \int\limits_{\mathcal{S}_f - \mathcal{S}_i}  d\hat{\mathcal{S}} \, \left[ \mathcal{E} - \mathcal{P}^a v_a \right] = \int\limits_{\mathcal{S}_f - \mathcal{S}_i}  d\hat{\mathcal{S}} \, \left[ \mathcal{E}_{\mathtt mat} + \mathcal{O} (\epsilon^2) \right],
\end{align}
where $\mathcal{E}_{\mathtt mat}$ is given in equation~(\ref{EMatter}) and the vacuum contributions on $\mathcal{S}_i$ and $\mathcal{S}_f$ cancel out. Thus, to the order we are working, we cannot use the left-hand side of the energy conservation law to determine $\Delta\mathrm{E}_{\mathtt{RQF}}$ at order $\epsilon^2$, e.g., changes in the Newtonian kinetic energy of masses in motion inside the RQF. However, we can obtain such $\epsilon^2$ information from the right-hand side, as we shall now see.

On the right-hand side of equation~(\ref{EnergyGCLPN}), the first term represents the matter energy flux and the second the gravitational energy flux. The former can be used to compute $\Delta\mathrm{E}_{\mathtt{RQF}}$ when, e.g., a particle enters or leaves the RQF sphere. However, since our primary interest is the gravitational energy flux we will set the matter energy flux to zero (i.e., $ T^{ab} n_a u_b = 0$) at the RQF surface. This leaves just the gravitational energy flux term, $- \alpha_a \mathcal{P}^a $. The presence of the lapse function in equation~(\ref{EnergyGCLPN}) accounts for a possible inhomogeneous time dilation across the system boundary as discussed in references~\cite{EMM2012,EMM2011}, but since the quasilocal momentum density is known only at order $\epsilon$ in our post-Newtonian approximation, the lapse function can be ignored. Hence, the (outward) gravitational energy flux is represented by
\begin{align} \label{adotPPN}
\alpha_i \mathcal{P}^i = \frac{\mathbb{S}^{ij}}{ c^4 \kappa r^2} (\mathbb{D}_i \Phi) \left( \frac{1}{2} \left( r \zeta_j \right)^\prime + \frac{1}{4} \mathbb{D}_j \left( r \zeta^\prime + \mathbb{D}^k \zeta_k \right)   \right) + \mathcal{O}(\epsilon^5).
\end{align}
In this form it is difficult to argue that this is what one should expect for the flux of gravitational energy.  However, it becomes clear that this is a sensible result by using our GEM fields,
equations~(\ref{GravitoelectricPN}) and ({\ref{GravitomagneticPN}), to calculate the GEM Poynting flux normal to the surface of the RQF sphere for comparison:
\begin{align} \label{GEMtoadotP}
r^I S^{\mathtt{GEM}}_I  = \frac{1}{c^3 \kappa} r^I \epsilon_{I}^{\phantom{I}JK} E^{\mathtt{GEM}}_J B^{\mathtt{GEM}}_K = \frac{1}{c^3 \kappa} \mathbb{E}^{ij} \, e^{\mathtt{GEM}}_i \, b^{\mathtt{GEM}}_j = \alpha_i \mathcal{P}^i + \mathcal{O}(\epsilon^5).
\end{align}
It is satisfying to see that, at leading order, our gravitational energy flux is really just the radial component of the GEM Poynting flux. It should be pointed out however that, while this is a useful tool for qualitatively understanding a cumbersome expression like equation~(\ref{adotPPN}), the GEM analogy quickly breaks down as a means of quantifying the flow of gravitational energy beyond leading order, and one should not hope to satisfy equation~(\ref{GEMtoadotP}) at higher orders \cite{EMM2011,Mashhoon}. On the other hand, based on the generality of the arguments in the previous section when constructing the conservation laws, $\alpha_a \mathcal{P}^a$ {\it will} continue to capture gravitational energy flow accurately at higher orders and so must, in fact, be {\it exactly} the gravitational Poynting vector.

In summary, as $\Delta t=t_f -t_i\rightarrow 0$, the completely general RQF energy conservation law in equation~(\ref{EnergyGCLPN}) reduces, in our post-Newtonian approximation, to
\begin{align}\label{EnergyRate}
\boxed{\frac{d\mathrm{E}_{\mathtt{RQF}}}{dt} =  \frac{1}{c^4 \kappa} \int \limits_{\mathcal{S}_t}  d\Omega\, \Phi \bigg[ \frac{1}{2} (r \mathbb{D}^k \zeta_k)^\prime + \frac{1}{4} \mathbb{D}^2 \left( r \zeta^\prime + \mathbb{D}^k \zeta_k \right)  \bigg] + \mathcal{O}(\epsilon^5)},
\end{align}
where we have used equation~(\ref{adotPPN}) and integrated by parts, and $d\Omega = \sin \theta \, d\theta \, d\phi$ is the surface element on a unit round sphere in standard spherical coordinates.

\subsubsection*{Linear Momentum}

To obtain a linear momentum conservation law we choose $\phi^a$ in equation~(\ref{MomentumGCLPN}) to be a boost CKV, which in the RQF coordinate system means taking $\phi^i = \frac{1}{r} \mathbb{B}^{Ii}$ . Here $I=1,2,3$ corresponds to a boost in the $X^I$ direction. The left-hand side of equation~(\ref{MomentumGCLPN}) then gives the change in the corresponding $\ell=1$ spherical harmonic component of the linear momentum inside the RQF between the surfaces of simultaneity $\mathcal{S}_i$ and $\mathcal{S}_f$, including the term $\frac{1}{c^2}\mathcal{S}^{ab} \phi_a v_b $ required to adjust for the motion of the RQF observers relative to $\mathcal{S}_i$ and $\mathcal{S}_f$:
\begin{align} \label{RQFMomentumDef}
\Delta\mathrm{P}^I_{\mathtt{RQF}} = \int\limits_{\mathcal{S}_f - \mathcal{S}_i}  d\hat{\mathcal{S}} \left[   \frac{\mathbb{B}^{Ii}}{r} \left( \mathcal{P}_i + \frac{1}{c^2}\mathcal{S}_{ij} v^j \right) \right].
\end{align}
Recall from equation~(\ref{QuasilocalMomentum}) that $\mathcal{P}_i$ (which is due to frame-dragging) begins at order $\epsilon$. Based on what we saw in the energy case, one might expect the relativistic stress term $\frac{1}{c^2}\mathcal{S}_{ij} v^j$ to be higher order, and thus negligible in our post-Newtonian approximation, but this turns out not to be the case because the quasilocal stress (in particular, the pressure) has a leading order vacuum term at order $\epsilon^{-2}$,
\begin{align}\label{QuasilocalStress}
\mathcal{S}_{ij} = - \frac{r}{\kappa} \mathbb{S}_{ij} - \frac{r}{c^2 \kappa} \bigg[ \big( \mathbb{D}_{(i} \mathbb{D}_{j)} - \mathbb{S}_{ij} \mathbb{D}^2 \big) \Phi \bigg] + \mathcal{O}(\epsilon^2),
\end{align}
while $v^j$ is of order $\epsilon^3$ (recall that $v_i = - u_i$). Therefore, the relativistic stress term actually contributes at the same order as $\mathcal{P}_i$ to the left-hand side of
equation~(\ref{MomentumGCLPN})---in general, both pieces are needed to account for the linear momentum measured by the RQF observers. Evaluating the integrand in equation~(\ref{RQFMomentumDef}) and integrating over $\mathcal{S}_t$ determines the linear momentum inside the RQF at time $t$:
\begin{align}\label{PNMomentumLHS}
\mathrm{P}^I_{\mathtt{RQF}} = \frac{r}{c^4 \kappa} \int\limits_{\mathcal{S}_t}  d\Omega \left[ \frac{r}{2} ( \zeta^{I})^\prime + \mathbb{B}^{I}_i \zeta^i - \frac{2 c^3}{r} r^I \on{3}{f^0} \right] + \mathcal{O}(\epsilon^3)
\end{align}
As a quick check of this equation, we imagine a Newtonian particle of mass $M$ moving with constant velocity $V$ through an RQF sphere. In the simplest case that we choose $\on{3}{f^0}$ in equation (\ref{VelocityPN}) such that $u_j =0$, it is easy to show that (the appropriate component of) $\mathrm{P}^I_{\mathtt{RQF}}$ equals $MV$ precisely when the particle is inside the RQF, and zero when it is outside.

However, our primary interest is in the right-hand side of equation~(\ref{MomentumGCLPN}), which has four terms representing fluxes of linear momentum. Apart from the matter linear momentum flux term, which we will turn off at the boundary of the RQF, it turns out that the dominant gravitational linear momentum flux term is the one involving the quasilocal pressure, $- {\rm P} \hat{D}_a \phi^a $.  From equation~(\ref{QuasilocalStress}) we find that this pressure is
\begin{align}
\mathrm{P} = \frac{1}{2}\sigma^{ij}\mathcal{S}_{ij} = - \frac{1}{\kappa r} + \frac{1}{2c^2 \kappa r} \mathbb{D}^2 \Phi + \mathcal{O}(\epsilon^2).
\end{align}
In our post-Newtonian approximation (which is not a full 1PN approximation---see footnote~\ref{1PN} on page~\pageref{1PN}), this pressure cannot be evaluated at order $\epsilon^2$ or higher. This in turn renders any information about the other fluxes at order $\epsilon^2$ and higher inconsequential since we cannot construct a complete picture of all of the fluxes. Unfortunately, one does not encounter non-zero net fluxes below order $\epsilon^2$.  We can see this by noting that, after integrating over time, which decreases the order in $\epsilon$ by one, the right-hand side of equation~(\ref{MomentumGCLPN}) can be evaluated at best at orders $\epsilon^{-3}$ and $\epsilon^{-1}$.  However, from equation~(\ref{PNMomentumLHS}), we already know that the left-hand side vanishes at these orders. This means that the fluxes on the right-hand side of equation~(\ref{MomentumGCLPN}) must integrate identically to zero in our post-Newtonian approximation. It is a straightforward calculation to verify this; we omit the calculation for the sake of brevity.

In the future, we intend to evaluate the gravitational linear momentum flux terms working with a higher-order post-Newtonian approximation.\footnote{See footnote~\ref{1PN} on page~\pageref{1PN}.} However, our present approximation scheme is sufficient to evaluate the gravitational angular momentum flux terms, to which we now turn our attention.

\subsubsection*{Angular Momentum}

To obtain an angular momentum conservation law we choose $\phi^a$ in equation~(\ref{MomentumGCLPN}) to be a rotation CKV, which in the RQF coordinate system means taking
$\phi^i = \mathbb{R}^{Ii} = \mathbb{E}^{i}_{\phantom{i}j}\mathbb{B}^{Ij}$. Here $I=1,2,3$ corresponds to a rotation about the $X^I$ axis. The left-hand side of equation~(\ref{MomentumGCLPN}) then gives the change in the corresponding $\ell=1$ spherical harmonic component of the angular momentum inside the RQF between the surfaces of simultaneity $\mathcal{S}_i$ and $\mathcal{S}_f$, including, as in the linear momentum case, the relativistic stress term:
\begin{align}\label{AngularMomentumPNCLLHS}
\Delta\mathrm{J}^I_{\mathtt{RQF}} = \int\limits_{\mathcal{S}_f - \mathcal{S}_i}  d\hat{\mathcal{S}}\, \left[   \mathbb{R}^{Ii} \left( \mathcal{P}_i + \frac{1}{c^2}\mathcal{S}_{ij} v^j \right) \right].
\end{align}
For the same reason as in the linear momentum case, we will, in general, need both terms in the integrand to compute the change in angular momentum inside the RQF.

On the right-hand side of equation~(\ref{MomentumGCLPN}), notice that, since the divergence of a rotation CKV is zero, the previously dominant flux, $- {\rm P} \hat{D}_a \phi^a $, is identically zero here and thus the angular momentum conservation law will contain more physics at this post-Newtonian order than the linear momentum law above. Taking the matter angular momentum flux term $-T^{ab} n_a \phi_b$ to be zero at the surface of the RQF leaves just two flux terms on the right-hand side of the conservation law. The first is the gravitational angular momentum flux $-\frac{1}{c^2} \mathcal{E} \alpha^a \phi_a$, which can be calculated at orders unity and $\epsilon^2$, while the other flux, $2 \nu \epsilon^{ab} \phi_a \mathcal{P}_b$, represents a Coriolis effect that starts at order $\epsilon^4$, and so can be neglected. Hence, the RQF angular momentum conservation law in our post-Newtonian approximation reduces to
\begin{align}\label{AngularMomentumPNCL}
\frac{d\mathrm{J}^I_{\mathtt{RQF}}}{dt} = - & \int \limits_{\mathcal{S}_t} d\hat{\mathcal{S}} \, \left[ N \frac{1}{c^2} \mathcal{E} \alpha^i \mathbb{R}^{I}_{i} + \mathcal{O}(\epsilon^4) \right].
\end{align}
The (outward) gravitational angular momentum flux is found to be
\begin{align} \label{AngularMomentumFlux}
N \frac{1}{c^2} \mathcal{E} \alpha^i \mathbb{R}^I_i = \mathbb{D}_i \bigg[ \frac{\mathcal{E}_{\mathtt{vac}}}{c^2} \mathbb{R}^{Ii} \big( \Phi +  \frac{1}{2c^2}(r\Phi^2)^\prime + \frac{1}{c^2}\Psi + \frac{1}{c} \partial_{t} (\Phi \on{1}{f^0}) \big) \bigg] + R^{Ii} \bigg[ \frac{r}{c^4} \mathcal{E}_{\mathtt{vac}} \dot{\zeta}_i + \frac{1}{c^2} \mathcal{E}_{\mathtt{mat}} \mathbb{D}_i \Phi \bigg] + \mathcal{O}(\epsilon^4).
\end{align}
The first group of terms in square brackets involves contributions at orders unity and $\epsilon^2$;  being a divergence, they will integrate to zero in equation (\ref{AngularMomentumPNCL}). Interestingly, this means that $\Psi$ does not show up in any of our post-Newtonian conservation laws, despite being necessary to compute them. This is an example of needing to work at a higher order during the intermediate steps of a perturbative calculation than is achieved in a final answer. On page 78 of reference~\cite{Wald1984}, Wald explains that in order to compute the acceleration of a test mass in linearized gravity, one makes use of the geodesic equation which is actually trivial in linearized gravity. The lesson is that it is standard, when analyzing Einstein's equation by perturbing around a flat background, that to find results at a particular order you may have to do certain elements of the calculation at a higher order. Also notice that the arbitrary time re-foliation parameter, $\on{1}{f^0}$, will thus not appear in the final result, leaving the integrated flux gauge-invariant. The physically relevant fluxes are thus contained in the second set of square brackets. Substituting the flux in equation~(\ref{AngularMomentumFlux}) into equation~(\ref{AngularMomentumPNCL}) and integrating by parts then gives the rate of change of angular momentum inside the RQF:
\begin{align}\label{AngularMomentumRate}
\boxed{\frac{d\mathrm{J}^I_{\mathtt{RQF}}}{dt} = \frac{r^2}{c^4 \kappa } \int  d\Omega \,\, \mathbb{R}^{Ii} \, \bigg[ 2 \dot{\zeta}_i + \Phi \, \mathbb{D}_i \left( 2 \Phi^\prime + \frac{1}{r} \mathbb{D}^2 \Phi \right) \bigg] + \mathcal{O}(\epsilon^4)}.
\end{align}

Equations~(\ref{EnergyRate}) and (\ref{AngularMomentumRate}) for the rates of change of energy and angular momentum inside an RQF in the lowest post-Newtonian approximation constitute the main result of this paper.  Given a metric in standard post-Newtonian form, equation~(\ref{WeinbergMetric}), one can use these relations to immediately compute the rate of change of energy and angular momentum inside an RQF due to gravitational fluxes passing through the RQF. Note that these rates are independent of the choice of time-foliation (i.e., these equations hold for arbitrary choice of the functions $\on{n}{f^0}$). In order to appreciate the utility of these equations let us now use them to analyze tidally interacting systems.

\section{Application to Tidal Interactions}\label{secTidalExamples}

Tidal interactions have acted as a test bed for analyzing conservation laws in general relativity by many authors in the past few decades; a good example is Hartle and Thorne~\cite{HT1985}.  In this section, we will first demonstrate the validity of the RQF approach by reproducing standard results for describing the transfer of energy and angular momentum via tidal interactions. The main technical advantage is that, unlike traditional methods, the RQF approach does {\it not} rely on pseudotensors---it is a manifestly geometrical (covariant) approach. Another key advantage is that it provides an entirely new way of understanding these interactions at a fundamental level, one that involves an exact realization of the equivalence principle in general relativity. We will then show that these equations have straightforward and practical applications by looking at two examples of tidal interactions in the solar system; in particular, the tidal heating of Jupiter's satellite Io and the mining of Earth's angular momentum by the Moon.

We begin with a pseudo-Cartesian coordinate system $X^A = (X^0=cT, X^I)$, $I=1,2,3$, as we used for the post-Newtonian metric. We then construct the metric describing a body whose center of mass is at rest at the spatial coordinate origin, immersed in the gravitational field of an arbitrary external body. We can characterize the gravitational field of the internal (central) body with a typical multipole moment expansion (see reference~\cite{Weinberg} for example), where we denote its mass by $M$, quadrupole moment by $Q_{IJ}$, angular momentum by $J_I$, and angular momentum current by $K_{IJ}$. The external gravitational field is described by the electric and magnetic parts of the Weyl tensor: $E_{IJ} = C_{0 I 0 J} $ and $B_{IJ} = \frac{1}{2} \epsilon_{I}^{\phantom{I}KL} C_{0JKL}$, respectively. In the de Donder gauge, the metric then takes the form of equation~(\ref{WeinbergMetric}) with parameters \cite{Zhang}
\begin{align}\label{PhiZeta}
\Phi &= - \frac{GM}{r} -\frac{3 G}{2 r^3} Q_{KL} r^K r^L +\frac{c^2 }{2} r^2  E_{KL} r^K r^L, \nonumber \\
\zeta_J &= -\frac{2G}{r^2} \epsilon_{JKL} J^K r^L - \frac{4G}{r^3} \epsilon_{JKL} K^K_{\phantom{K}M} r^L r^M - \frac{2 c^3 }{3} r^2 \epsilon_{JKL} B^K_{\phantom{K}M} r^L r^M \nonumber \\ &\quad\, -\frac{2G}{r^2} \dot{Q}_{JK} r^K - \frac{10}{21} c^2 r^3 \dot{E}_{KL} r_J r^K r^L + \frac{4}{21} c^2 r^3 \dot{E}_{JK} r^K.
\end{align}
Note that all of the rank-two tensors in equation~(\ref{PhiZeta}) are symmetric and trace-free. Furthermore, the internal quadrupole moment is defined with the convention $Q_{IJ} = \int d^3 X \, \rho \left( X^I X^J - \frac{1}{3} R^2 \delta^{IJ} \right)$.

In this spacetime, we now embed an RQF enclosing and centered on the internal body, but not enclosing the external body (i.e., $ \mathcal{L} \ll r \ll \mathcal{R}$ where $\mathcal{L} \sim GM/c^2$ is the gravitational length scale associated with the internal body and $\mathcal{R}$ is the radius of curvature of the external gravitational field, which is related to the Ricci scalar by $R \sim \mathcal{R}^{-2}$). As promised above, we can now simply substitute the metric functions in equation~(\ref{PhiZeta}) into the conservation laws in equations~(\ref{EnergyRate}) and (\ref{AngularMomentumRate}). Let us first look at the rate of change of energy.  Carrying out the angular integration we obtain
\begin{align}
\frac{d\mathrm{E}_{\mathtt{RQF}}}{dt} &= -\frac{c^2}{2} E_{IJ} \dot{Q}^{IJ} +  \frac{d}{dt}\big[ -\frac{1}{10} c^2 E_{IJ} Q^{IJ} +\frac{9}{10}\frac{G}{r^5} Q_{IJ} Q^{IJ} +\frac{1}{60}\frac{c^4}{G} r^5 E_{IJ} E^{IJ} \big] + \mathcal{O}(\epsilon^5). \label{TidalEnergyRate}
\end{align}
This equation represents the total rate of inward flow of gravitational field energy through the RQF, which in general fluctuates with time. We would like to determine which term or terms represent the rate of mechanical work done by the external field on the internal body---the tidal work. In the case of the Jupiter-Io system, most of this tidal work represents a transfer of thermal energy to Io, and so is sometimes also referred to as tidal heating~\cite{Purdue}. Regarding the time derivative of the terms in square brackets, the first represents flow of field interaction energy, the second, flow of self field energy, and the third, flow tidal field energy. Integrating these flows over time yields the changes in the amounts of these forms of gravitational energy stored inside the RQF, which can fluctuate as the internal body is deformed or the distance to the external body changes. Following Purdue~\cite{Purdue} we observe that these three types of energy can depend only on the instantaneous self and tidal fields, and so their flow rates must be total time derivatives. Moreover, if the tidal field changes but the self field does not (no tidal deformation), there is no tidal work done. Thus, the power flow due to tidal mechanical work must be just the first term on the right-hand side of equation (\ref{TidalEnergyRate}):
\begin{align}\label{TidalPower}
\boxed{P_{\mathtt{tidal}} = -\frac{c^2}{2} E_{IJ} \dot{Q}^{IJ}}.
\end{align}
It is this term that will be of interest in practical discussions regarding the transfer of thermal energy via tidal interactions as we will see in {\S}\ref{secsubSolarExamples} below.

It is worthwhile to compare this analysis to those based on the traditional pseudotensor approach. Purdue~\cite{Purdue}, for example, starts with a Newtonian analysis of tidal heating with a parameter $\alpha$ that characterizes different ways to localize the Newtonian gravitational energy. This results in a numerical coefficient $(2+\alpha)/10$ in front of the interaction energy term---see her equation~(20), which is $-1/10$ in our case ($\alpha=-3$). She then does a general relativistic analysis using the Landau-Lifshitz pseudotensor and the de Donder gauge, and arrives at a result that is similar to our equation~(\ref{TidalEnergyRate}), with $\alpha=-3$, but with different numerical coefficients for the $Q_{IJ} Q^{IJ}$ and $E_{IJ} E^{IJ}$ terms ($63/20$ and $-7/30$ versus our $9/10$ and $1/60$)\footnote{Purdue~\cite{Purdue} does not give these coefficients in her paper---we computed them by following her procedure.}. At the very least, it seems unlikely to us that these coefficients should have opposite signs; in particular, the $E_{IJ} E^{IJ}$ term should probably be positive. Moreover, she then considers the Landau-Lifshitz pseudotensor in a general gauge (coordinate transformation) and finds that the numerical coefficients for the $E_{IJ} Q^{IJ}$ and $E_{IJ} E^{IJ}$ terms are gauge-dependent, allowing the sign of the $E_{IJ} E^{IJ}$ term to be negative or positive. The point is that, although the tidal work term is gauge-invariant, a pseudotensor will give different answers for the other energy fluxes, depending on the choice of coordinate system. One expects energy fluxes to depend on the choice of observer (energy is an observer-dependent construct), but in a pseudotensor approach there is no simple geometrical or physical relationship between choice of observer and choice of coordinate system. In the RQF approach, the coordinate system is fixed by a simple, physically sensible geometrical condition, and we get what seem to be physically sensible results.

It is also worth pointing out that Booth and Creighton~\cite{Booth} have analyzed tidal heating by evaluating the Brown and York~\cite{BY1993} quasilocal energy in a general coordinate system, and our result corresponds to a particular choice of their general coordinate system. But again, in reference~\cite{Booth} no connection is made between the choice of coordinate system and the physical or geometrical properties of the observers. The RQF approach is able to identify {\it true} fluxes relative to observers in a geometrically well-defined and physically sensible (rigid) quasilocal frame.

Consider next how the angular momentum inside the RQF changes.  For the tidal metric above, equation (\ref{AngularMomentumRate}) becomes, after integration,
\begin{align}\label{TidalAngularMomentumRate}
\frac{d\mathrm{J}^I_{\mathtt{RQF}}}{dt} &=   c^2 \epsilon^{IJK} E_{JL} Q_K^{\phantom{K}L} - \frac{4}{3} \dot{J}^I + \mathcal{O}(\epsilon^4),
\end{align}
where we have made use of the identity $\mathbb{R}^J_j = \epsilon^J_{\phantom{J}KL} r^K \mathbb{B}^L_j$.  Analogous to the energy equation above, this equation gives the rate of change of the angular momentum inside the RQF. The first term on the right-hand side describes the physical torque associated with the tidal forces acting on the quadrupole moment of the internal body:
\begin{align}\label{TidalTorque}
\boxed{\tau^I_{\mathtt{tidal}} = c^2 \epsilon^{IJK} E_{JL} Q_K^{\phantom{K}L}}.
\end{align}
This is the important piece for physical applications.

The second term on the right-hand side of equation~(\ref{TidalAngularMomentumRate}), $-\frac{4}{3} \dot{J}^I$, has an interesting physical interpretation. As the angular momentum of the internal body {\it changes} (e.g., spin-down of the Earth in the Earth-Moon system), there is a {\it change} in the degree of frame-dragging of the surrounding space. In order to remain fixed relative to the distant stars, the RQF observers must then tangentially accelerate in the opposite direction: $\alpha_i = \frac{r}{c^2} \dot{\zeta}_i$ (see equation (\ref{AccelPN})). In previous work~\cite{EMM2012} we have suggested that the gravitational vacuum energy may be a {\it real} source of ``mass" associated with space itself. If so, the RQF observers are accelerating tangential to a surface with effective (negative) mass density $\rho_{\mathtt{vac}} = \frac{1}{c^2} \mathcal{E}_{\mathtt{vac}}$.  This angular acceleration leads to a change in the angular momentum inside the system in exactly the same way that accelerating linearly relative to an object in inertial motion changes the linear momentum that the accelerating observer ascribes to that object, at a rate equal to the mass of the object times the acceleration of the observer. If we look back at equation~(\ref{AngularMomentumFlux}), we can see that it is precisely this mechanism that is the origin of the flux term in question: $\rho_{\mathtt{vac}}\times (R^{Ii}\alpha_i)= R^{Ii} \frac{r}{c^4} \mathcal{E}_{\mathtt{vac}} \dot{\zeta}_i$. As a verification of this argument, we can use the $\beta^I (t,r)$ freedom discussed just before equation~(\ref{RQFSolution}) to put the RQF observers in a locally non-rotating frame, $\nu=0$ (i.e., the frame-dragged frame that rotates with respect to the distant stars), and show that this frame is locally inertial---that is, the RQF observers experience no angular acceleration in this frame, even when the angular momentum of the internal body {\it changes}. Rotating in this way {\it with} the gravitational vacuum, no acceleration means no change in angular momentum; in this frame, there is only the tidal torque term of equation~(\ref{TidalTorque}) on the right-hand side of equation~(\ref{TidalAngularMomentumRate}).

In order to compare our result to other work, it is important to make the distinction here between the angular momentum inside the RQF, $\mathrm{J}^I_{\mathtt{RQF}}$, and the angular momentum of the internal body, $J^I$.  In reference~\cite{HT1985}, Hartle and Thorne use a pseudotensor approach to derive the rate of change of the angular momentum of the internal body as solely arising from the tidal torque in
equation~(\ref{TidalTorque}): $\dot{J}^I = c^2 \epsilon^{IJK} E_{JL} Q_K^{\phantom{K}L}$.  We can now reproduce Hartle and Thorne's result by evaluating the left-hand side of the conservation law in equation~(\ref{AngularMomentumPNCLLHS}), independently of the right-hand side. Making use of equations~(\ref{QuasilocalMomentum}) and (\ref{QuasilocalStress}) we find
\begin{align}\label{AngularMomentumLHS}
\int\limits_{\mathcal{S}_t}  d\mathcal{S} \, \bigg[   \mathbb{R}^I_i \frac{d}{dt} \left( \mathcal{P}^i + \frac{1}{c^2}\mathcal{S}^{ij} v_j \right) \bigg]  = \frac{1}{2 c^4 \kappa} \int d\Omega \, \mathbb{R}^{Ii} \,\big( r^3  \dot{\zeta}_i \big)^\prime = - \frac{1}{3} \dot{J}^I + \mathcal{O}(\epsilon^4).
\end{align}
The factor of $-\frac{1}{3}$ can be understood by invoking the same gravitational vacuum energy argument we used for the right-hand side. The RQF observers see an angular momentum made up of two contributions: first, there is the angular momentum of the internal body, $\mathrm{J}^I_{\mathtt{body}} = J^I$; second, there is the angular momentum associated with rotating relative to the gravitational vacuum energy, $\mathrm{J}^I_{\mathtt{vac}} = -\frac{4}{3} J^I$.  Therefore, the total angular momentum that the RQF observers see is the sum of the two: $\mathrm{J}^I_{\mathtt{RQF}} = \mathrm{J}^I_{\mathtt{body}} + \mathrm{J}^I_{\mathtt{vac}} = -\frac{1}{3} J^I$.  Conveniently, we can now equate (\ref{TidalAngularMomentumRate}) and (\ref{AngularMomentumLHS}) to obtain
\begin{align}
\frac{d J^I}{dt} &=   c^2 \epsilon^{IJK} E_{JL} Q_K^{\phantom{K}L} + \mathcal{O}(\epsilon^4) \label{TidalAngularMomentumHT}.
\end{align}
in agreement with equation~(3.17b) in reference~\cite{HT1985}.\footnote{Note that in reference~\cite{HT1985}, Hartle and Thorne perform a slightly different expansion than we do.  In particular, they treat the gravitoelectric and gravitomagnetic fields to be formally of the same order.  The same is done for the mass quadrupole momentum and angular momentum current.  However, in the post-Newtonian approximation, the gravitomagnetic and angular momentum current are each an order in $\epsilon$ {\it smaller} than their counterparts.  As a result, a term of the form $\epsilon^{IJK} B_{JL} K_K^{\phantom{K}L}$ which also appears in their equation (6.23) would be of order $\epsilon^6$ in our expansion.  This is why it does not appear in equation~(\ref{TidalAngularMomentumHT}).}

Traditionally, the tidal interactions in equations~(\ref{TidalPower}) and (\ref{TidalTorque}) are explained in terms of the Newtonian gravitational force. Roughly speaking, the rate of tidal power in equation~(\ref{TidalPower}) has the form ${\bf F}\cdot {\bf v}$ (power), where the gravitational tidal force $E_{IJ}$ represents the force, $\bf F$, and the rate of change of the quadrupole moment $\dot{Q}^{IJ}$ represents the velocity, $\bf v$. Similarly, the tidal torque in equation~(\ref{TidalTorque}) has the form ${\bf r}\times {\bf F}$ (torque), where $E_{IJ}$ again represents the force, and ${Q}^{IJ}$ represents the position, ${\bf r}$ (of the mass(es) in the system). The RQF approach, on the other hand, reveals that the ultimate explanation of these interactions is not found in a gravitational force, but rather in an exact, quasilocal manifestation of the equivalence principle in general relativity. To see this, imagine being inside an accelerating box in flat spacetime that contains a freely-floating point particle of mass $m$ that appears to be accelerating toward you; the particle's kinetic energy and momentum (relative to you) increase due to the acceleration of your frame. In the context of special relativity, the proper time rates of change of the relativistic energy and momentum of the particle are $- a_a p^a$ and $-(E/c^2) a_a \phi^a$, respectively, where $a_a$ is your four-acceleration (at the moment the particle passes you), $E$ and $p^a$ are the instantaneous relativistic energy and four-momentum of the particle relative to you, and $\phi^a$ is a unit vector orthogonal to your four-velocity and representing the particular spatial component of momentum you are measuring. Since energy and momentum are conserved quantities we must ask: Where does the new energy and momentum in your box come from? In the context of special relativity the answer is: ``Nowhere---energy and momentum are frame-dependent constructs, and we are just changing the frame!"

But by the equivalence principle (and assuming for the moment constant proper acceleration, just for simplicity of the argument), we are to imagine that the accelerating box is physically equivalent to (experimentally indistinguishable from) a box at rest in a uniform gravitational field, and that the particle is experiencing an acceleration toward you due to the ``force" of gravity. This ``force" acting over time transfers energy and momentum to the particle. But where does the energy and momentum come from? If it was an electromagnetic force we would say the energy and momentum come from the electromagnetic field (and ultimately, the source of that field). Since the ``force" in our case is gravitational, the energy and momentum must come from the gravitational field (and ultimately, the source of that field). According to the equivalence principle, there must exist surface fluxes (at the boundary of the box) representing net gravitational energy and momentum entering the box from the outside. According to equation~(\ref{EnergyGCLPN}), the gravitational energy flux in question is precisely $-\alpha_a {\mathcal P}^a$, which has dimensions of energy per unit area per unit time; and according to equation~(\ref{MomentumGCLPN}), the gravitational momentum flux in question is $-({\mathcal E}/c^2)\alpha_a \phi^a$, which has dimensions of momentum per unit area per unit time.\footnote{For simplicity, in equation~(\ref{MomentumGCLPN}) we are assuming that the twist, $\nu$, of the RQF congruence vanishes. Also, in general, the pressure term, ${\rm P} \mathcal{D}_{a} \phi^{a}$ is part of the gravitational momentum flux, but in the case that $\phi^a$ is a rotational CKV, which is the case of interest for tidal torque, $\mathcal{D}_{a} \phi^{a}=0$.} These are the {\it exact} quasilocal general relativistic analogues of the local special relativistic expressions $- a_a p^a$ and $-(E/c^2) a_a \phi^a$, respectively.

In the RQF approach, $\alpha_a$ is the proper acceleration tangent to the boundary of the system that is required in order for the observers to maintain quasilocal rigidity. In the quasilocal spirit, the system inside the RQF is represented entirely in terms of the intrinsic and extrinsic geometry of the boundary, ${\mathcal B}$. As far as any measurements of what is inside are concerned, the system {\it is} the boundary. Using a gyroscope, each RQF observer measures the local momentum density ${\mathcal P}^a$ inside their small ``box" (an infinitesimal two-dimensional patch of the RQF sphere), notices their box (local frame) has an acceleration $\alpha_a$, and concludes that the proper time rate of change of the energy density inside their box is $-\alpha_a {\mathcal P}^a$, due simply to their acceleration relative to the existing momentum in their box. Multiplying by the area element and integrating over all observers, and then over proper time (between the two simultaneities $\mathcal{S}_i$ and $\mathcal{S}_f$), they get the change in the total energy inside the system. Similarly, using a ruler, each RQF observer measures the local energy density ${\mathcal E}$ inside their small box, notices their box has an acceleration $\alpha_a$, and concludes that the proper time rate of change of the $\phi^a$-component of the momentum density inside their box is $-({\mathcal E}/c^2)\alpha_a \phi^a$, due simply to their acceleration relative to the existing mass-energy in their box. In the case that $\phi^a$ is a {\it rotational} CKV, the observers are dealing with {\it angular} momentum. Again, integrating over all observers, and then over proper time, they obtain the change in the total angular momentum inside the system.

The RQF approach reveals that, apart from matter fluxes, there is only one mechanism for the transfer of energy-momentum across a system boundary: acceleration relative to the energy-momentum inside the system. This is the simplest possible mechanism in Newtonian mechanics, and it turns out to be universal, applying even in the strong field regime of general relativity. The extension of this mechanism from Newtonian mechanics to general relativity requires a shift from local to quasilocal: both the acceleration and energy-momentum information are encoded quasilocally in the intrinsic and extrinsic curvature of the boundary. We claim that this is an exact, quasilocal manifestation of the equivalence principle in general relativity. In the first post-Newtonian approximation, $\alpha_a$, ${\mathcal E}$, and ${\mathcal P}^a$ will depend, of course, on the quadrupole moment and the tidal field, and so the exact general relativistic energy and momentum fluxes will reduce to products of these, which integrate to equations~(\ref{TidalPower}) and (\ref{TidalTorque}), or more generally, equations~(\ref{TidalEnergyRate}) and (\ref{TidalAngularMomentumRate}). These approximate expressions can be interpreted in terms of a Newtonian gravitational force, but their true origin has nothing to do with a gravitational force, and everything to do with the equivalence principle. The exact general relativistic explanation of energy-momentum transfer, in tidal interactions and more generally, is fundamentally different than in Newtonian gravity: it is simpler, and it is non-local (quasilocal).

\subsection{Examples: Solar System Dynamics} \label{secsubSolarExamples}

We will now look at two examples  within our solar system to test the utility of the results above.  Specifically, we want to test the formulas for tidal power, equation~(\ref{TidalPower}), and tidal torque, equation~(\ref{TidalTorque}). It is well-known \cite{Lopes} that Jupiter's satellite Io is volcanically active and that this activity cannot be explained without taking into account the enormous time-varying tidal forces exerted on Io in its eccentric orbit around Jupiter. This scenario is an ideal one to apply the equation for tidal power to quantify the rate of energy transfer and compare with observation. Another well-documented phenomenon is the recession of the Moon in its orbit around the Earth~\cite{StaceyDavis}. This is because the tidal field of the Moon creates tidal bulges on the Earth, but the rotation of the Earth causes these bulges to rotate ahead of the common axis joining the two bodies. In turn, the Moon then pulls on the forward (closer) bulge more than it does on the trailing (farther) bulge.  The resulting net torque, which we will calculate using our results above, transfers spin angular momentum of the Earth to orbital angular momentum of the Earth-Moon system resulting in an increase in the orbital distance.
	
In both of these examples it will be useful to employ a result of Love's \cite{Love} which will allow us to relate the quadrupole potential of the internal body, $\Phi_{\mathtt{quad}}$, to the the tidal potential of the external body, $\Phi_{\mathtt{tidal}}$.  The basic idea is that the quadrupole moment exists only because the distorted shape of the internal body is due to the external body's tidal potential. As such, one should be able to relate the two potentials by a numerical factor, $k_2$, called the Love number, which characterizes how easily the internal body is deformed. Specifically, the potentials should satisfy the relation~\cite{StaceyDavis}
\begin{align}\label{LoveRelationBasic}
\Phi_{\mathtt{quad}}(t,\psi) = k_2 \frac{R_{\mathtt{int}}^5}{r^5} \Phi_{\mathtt{tidal}}(t-\tau,\psi-\delta)
\end{align}
where $R_{\mathtt{int}}$ is the radius of the internal body, $\psi$ is the angle between the common axis joining the two bodies and a point in space, $\delta$ is the angle that the quadrupole is carried ahead of the common axis due to the rotation of the internal body, and $\tau$ is the lag in the tide due to the finite time that it takes the internal body to deform. For a tidally locked satellite like Io, the absence of rotation means that $\delta = 0$. For the Earth-Moon system, the tidal forces on the Earth do not fluctuate significantly with time, so the shape of the Earth is constant in time. Thus, $\tau =0$---indeed the time-dependence in equation~(\ref{LoveRelationBasic}) can be ignored altogether.

It will be useful to recast equation~(\ref{LoveRelationBasic}) in terms of the notation used in this paper---that is, in terms of the quadrupole moment tensor, $Q_{IJ}$, and external tidal field tensor, $E_{IJ}$.  If we momentarily define new spatial coordinates $\bar{X}^{\bar{I}}$ that are related to the standard $X^I$ by a rotation through an angle $\delta$ ahead of the common axis between the two bodies then we should be able to write $\bar{Q}_{\bar{I}\bar{J}}(t) = - \lambda k_2 E_{IJ}(t-\tau)$ where $\lambda$ is a dimensionful constant.  If we substitute this relation into the metric function for the full potential $\Phi$ in equation (\ref{PhiZeta}) it is straightforward to show that, in order to satisfy Love's original relation (\ref{LoveRelationBasic}), we must take $\lambda = \frac{1}{3} \frac{c^2}{G} R_{\mathtt{int}}^5$ which leads to the relation\footnote{Note that this relation differs from recent work on tidal effects in neutron stars (see references \cite{Damour} and \cite{Poisson} for example).  In these references, the term `Love number' is used to refer to the apsidal constant which actually differs from the standard Love number by a factor of two, $k_{2,\mathtt{Love}} = 2k_{2,\mathtt{apsidal}}$.  Here we choose to maintain Love's original definition of the Love number and that used by the geophysics community where one typically has to turn to find values of the Love number for bodies in our solar system.}
\begin{align}\label{LoveRelation}
\bar{Q}_{\bar{I}\bar{J}}(t) = - \frac{1}{3} \frac{c^2}{G} R_{\mathtt{int}}^5 k_2 E_{IJ}(t-\tau).
\end{align}

\subsubsection*{Tidal Power in the Jupiter-Io System}

\begin{figure}
\begin{center}
\includegraphics[scale=0.5]{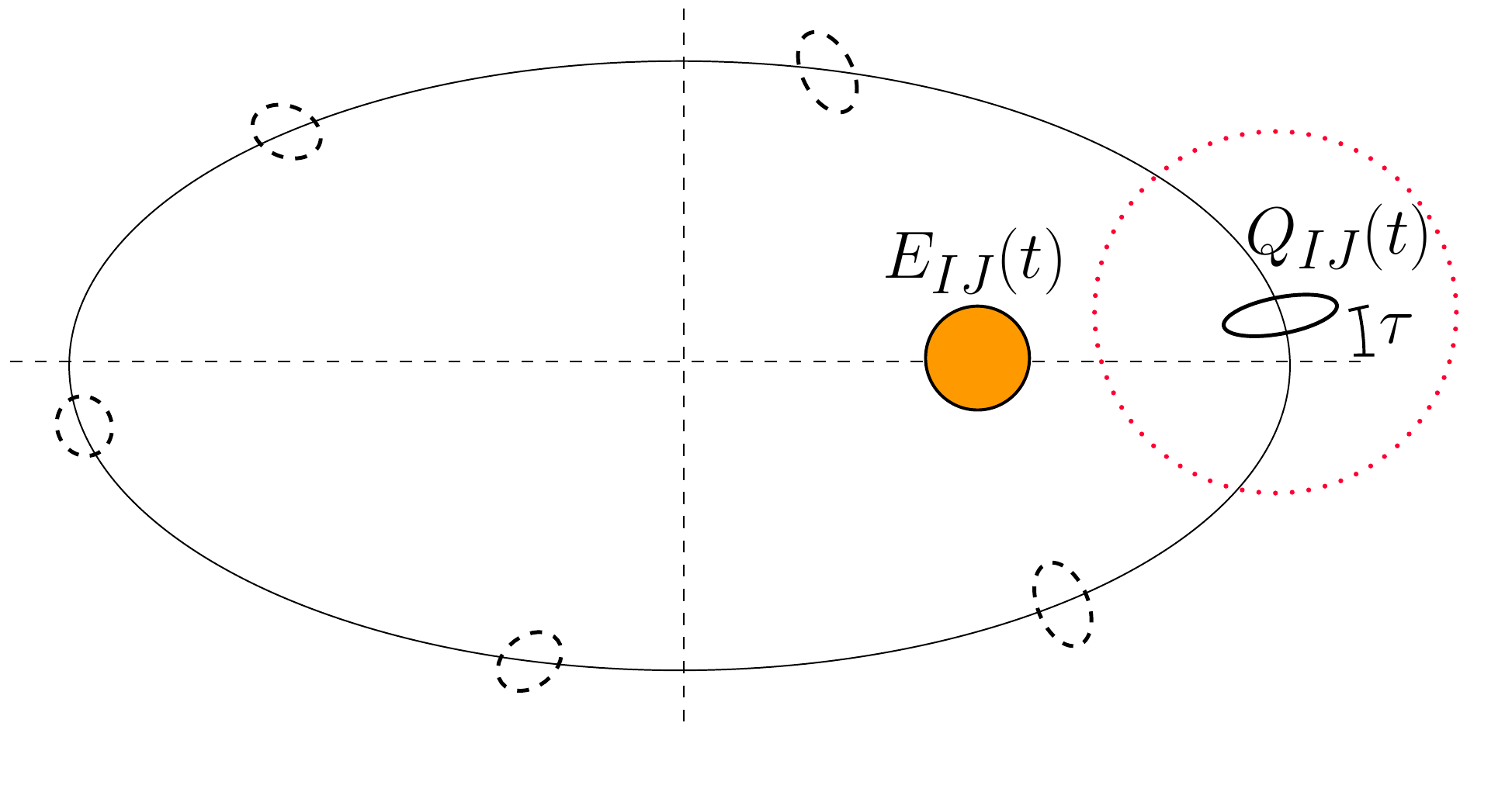}
\caption{The tidal field of Jupiter, $E_{IJ}$, induces a quadrupole moment, $Q_{IJ}$, in Io which is delayed by a lag time $\tau$ due to the finite time it takes for the deformation to fully set in.  Furthermore, the elliptical nature of the orbit means that the strength of Jupiter's tidal field varies at Io and, in turn, the degree to which Io gets distored varies too.  This results in a continuous transfer of gravitational energy from Jupiter to Io which, on time averaged, is dissipated mostly thermally. By placing an RQF around Io and averaging equation (\ref{TidalPower}) over one full orbit, we determine the average power of this tidal heating.}\label{JupiterIo}
\end{center}
\end{figure}

Let us now put an RQF around Jupiter's moon Io and use equation (\ref{TidalPower}) to compute the work done on Io during one orbit.  As mentioned above, Io is tidally locked to Jupiter ($\delta = 0$) so the coordinates $\bar{X}^I$ are just the $X^I$ coordinates.  However, we {\it do} need to take into account that the tidal bulge from the external field at a given time induced on Io does not occur instantaneously---there is a time lag, $\tau$, in the amplitude of the bulge (see figure~\ref{JupiterIo}). The tidal work per unit time is then
\begin{equation}\label{TidalPowerIo}
\boxed{P_{\mathtt{tidal}}  = -\frac{c^2}{2} E_{IJ}(t) \dot{Q}^{IJ}(t) = \frac{c^4}{4G} k_{2,\mathtt{Io}} R_{\mathtt{Io}}^5 E_{RR}(t) \dot{E}_{RR}(t-\tau)},
\end{equation}
where we have made use of equation (\ref{LoveRelation}).  Here, $E_{RR}(t) = -\frac{2G M_J}{c^2 R(t)^3}$ is the radial-radial component of the tidal field of Jupiter.  The lag time, $\tau$, is unfortunately not well known but, according to \cite{Yoder}, should go like the period of the orbit divided by the dissipation factor, $Q$, which is typically assumed to have the value $Q \simeq 100$ for Io \cite{MurrayDermott}.  This means $\tau$ should be approximately $\tau \simeq 25$ minutes for Io.  The Love number for Io is also not well known because its calculation relies on knowing the rigidity, $\mu$, of Io.  The usual way to get around this is to assume that Io has the rigidity of a typical rocky body, $\mu \simeq 5 \times 10^{10} \, Pa$.  This yields an approximate Love number for Io of $k_{2,\mathtt{Io}} \simeq 0.03$ \cite{MurrayDermott}.  Using these parameters, we then numerically integrate the power over one full orbit and divide by the orbital period to find the average power transferred to Io.  This yields an average power of $\langle P_{\mathtt{tidal}} \rangle \simeq 1.6 \times 10^{14} \, W$.  The currently accepted value based on various models and observations for the heat generated through tidal interactions in Io is $0.6-1.6 \times 10^{14} \, W$ \cite{Lopes}.   Our value agrees well with observation and we expect that with better knowledge of the Love number, tidal lag time, and a more accurate description of the time dependence of the quadrupole moment our set up could be used to compute even more accurately the actual amount of tidal heating in Io. The main lesson from this exercise, however, is that equation~(\ref{TidalPower}) {\it is} the correct expression for tidal power, derived here using a covariant approach, and it has utility in real-world problems like the one above.

\subsubsection*{Tidal Torque in the Earth-Moon System}

\begin{figure}
\begin{center}
\includegraphics[scale=0.7]{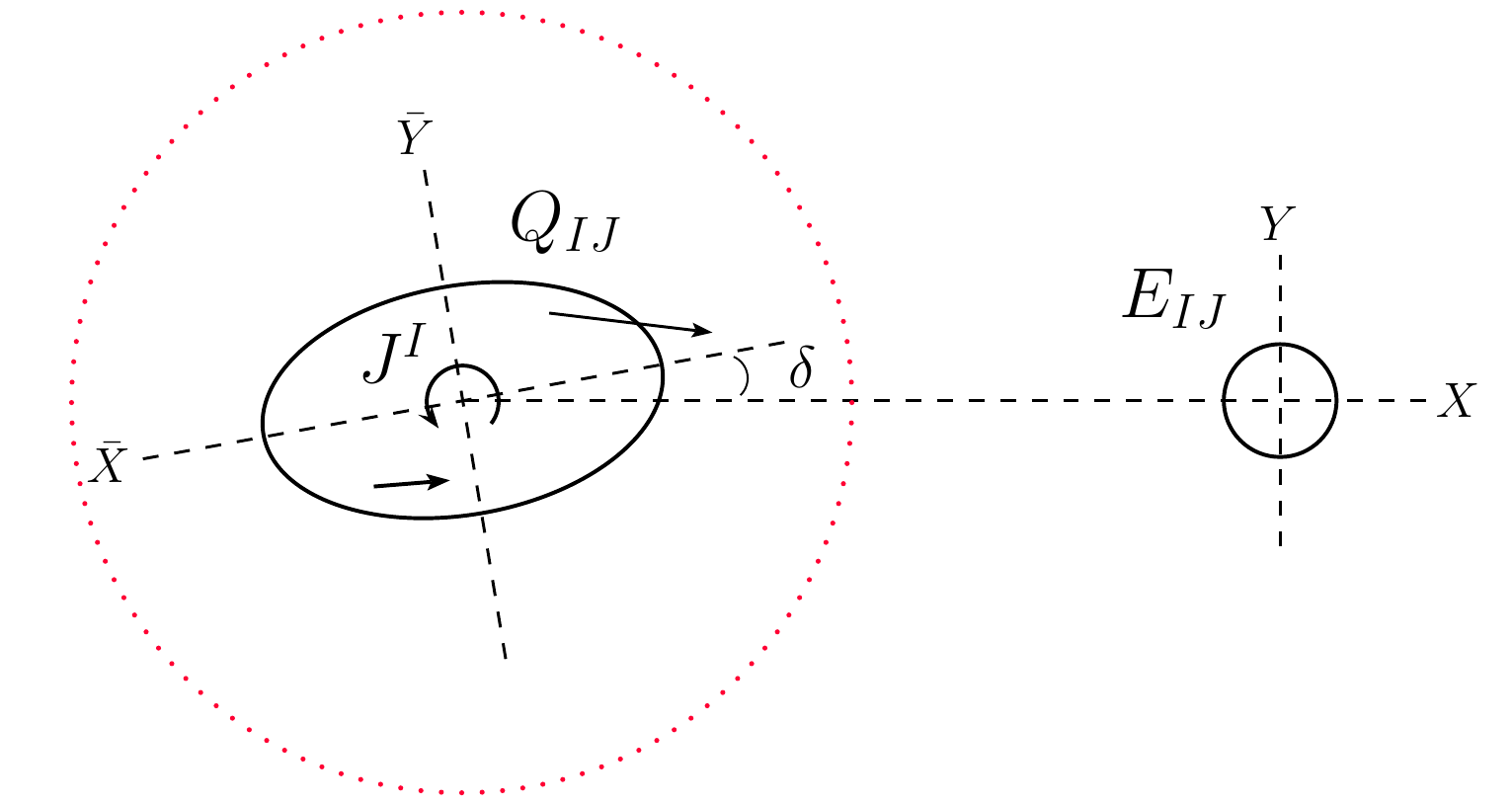}
\caption{The Moon's gravitational field induces a tidal bulge on the Earth, $Q_{IJ}$, which is carried ahead of the Earth-Moon axis by an angle $\delta$ due to the rotation of the Earth, $J^I$.  The tidal field of the Moon, $E_{IJ}$, then pulls on the near-side bulge more strongly than the far-side bulge which results in a net torque slowing down the rotation of the Earth.  Angular momentum is thus transferred from the spin of the Earth to the Earth-Moon orbit resulting in the recession of the Moon. We analyze this angular momentum transfer by centering an RQF around the Earth and computing the instantaneous torque using
equation~(\ref{TidalTorque}).}\label{EarthMoon}
\end{center}
\end{figure}

In our next example, we consider the Earth-Moon system and center our RQF around the Earth, such that the RQF does not rotate relative to the line connecting the centers of the Earth and Moon (see figure~\ref{EarthMoon}). This means that the RQF rotates relative to the distant stars (the zeroth order post-Newtonian spacetime), but this effect can be ignored in the analysis (a similar comment applies to the Io calculation above), as can the wobble of the center of the Earth relative to the center of mass of the Earth-Moon system.  Since the Earth-Moon orbit has negligible eccentricity, the quadrupole moment and tidal field do not vary appreciably during the orbit.  However, the Earth's tidal bulge {\it does} rotate ahead of the Moon and, as a result, it carries its bulge ahead of their common axis by an angle $\delta$. For concreteness, let us take the Earth-Moon system to be connected by the $X$-axis, with the $Z$-axis parallel to the rotational axis of the Earth. We will then be interested in computing the rate at which the $Z$-component of the angular momentum of the Earth, $J^Z$, changes. This is just given by equation~ (\ref{TidalTorque}) for the tidal torque.  We find that
\begin{align}\label{TidalTorqueEarthStep1}
\tau^Z_{\mathtt{tidal}} = \frac{3}{2} c^2 E_{XX} Q_{XY}
\end{align}
where we have used the fact that $E_{IJ}$ is diagonal and $E_{YY} = -\frac{1}{2} E_{XX}$.

To relate the quadrupole moment to the tidal field tensor we first   need to work out how $Q_{XY}$ is related to the components of $\bar{Q}_{\bar{I}\bar{J}}$.  The transformation between $X^I$ and $\bar{X}^{\bar{I}}$ coordinates is simply a rotation about the $Z$-axis by an angle $\delta$.  Specifically, $X = \bar{X} \cos \delta - \bar{Y} \sin \delta$ and $Y = \bar{Y} \cos \delta + \bar{X} \sin \delta$.  Therefore, a quick calculation leads to $Q_{XY} = \frac{3}{4} \bar{Q}_{\bar{X}\bar{X}} \sin (2 \delta)$.  Now we can use equation~(\ref{LoveRelation}) which relates $\bar{Q}_{\bar{X}\bar{X}}$ to $E_{XX}$, the tidal field of the Moon at Earth, to show
\begin{align}\label{TidalTorqueEarth}
\boxed{\tau^Z_{\mathtt{tidal}} = -\frac{3}{2} k_{2,\mathtt{E}} G M_M^2 \frac{R_E^5}{R_{EM}^6} \sin (2\delta)}.
\end{align}
This reproduces the Newtonian result exactly (see equation~(8.20) in reference~\cite{StaceyDavis}) and, using the measured values of $\delta = 2.89^\circ$ and $k_{2,\mathtt{E}} = 0.245$, along with standard values for all other parameters, gives a net torque of $4.4 \times 10^{16}$ kg m${}^2$ s${}^{-2}$.  As discussed above, the back reaction of this torque transfers angular momentum to the orbital motion of the Earth-Moon system and results in a recession rate for the Moon of $37$ millimetres per year, which agrees precisely with the observed value~\cite{StaceyDavis}. Again, the main lesson from this exercise is that equation~(\ref{TidalTorque}) is the correct expression for tidal torque, derived here using a covariant approach, and it has important real-world applications.

\section{Summary and Conclusions}\label{secSummary}

In this paper we reviewed the rigid quasilocal frame (RQF) approach to constructing completely general energy, linear momentum and angular momentum conservation laws for extended systems in general relativity. We derived the general form of these laws in a first post-Newtonian (1PN) approximation\footnote{See footnote~\ref{1PN} on page~\pageref{1PN}.}, illustrating explicitly the 1PN approximation of the exact RQF gravitational energy and angular momentum fluxes---see equations~(\ref{EnergyRate}) and (\ref{AngularMomentumRate}). We also provided an elegant description of the energy conservation law in terms of the gravitoelectromagnetic (GEM) analogy: at lowest order, the gravitational energy flux that the RQF observers measure is, indeed, the radial component of the GEM Poynting vector.

We then applied these 1PN conservation laws to analyze gravitational tidal interactions. We derived the generally accepted tidal power and tidal torque formulas---see equations~(\ref{TidalPower}) and (\ref{TidalTorque}). The two new aspects of our results are: (1) we derived these formulas using a purely geometrical (manifestly covariant) analysis rather than one based on a stress-energy-momentum pseudotensor, and (2) we applied these formulas to two concrete examples (the Jupiter-Io and Earth-Moon systems), providing explicit formulas in terms of Love numbers---see equations~(\ref{TidalPowerIo}) and (\ref{TidalTorqueEarth}). Putting in numbers, the theoretical predictions match closely with the observations. Note that these formulas are usually derived in the context of Newtonian gravity; our presentation shows how one can arrive at them in the context of general relativity, giving an explicit explanation of the Love relation between the quadrupole and tidal field tensors, which does not seem to be readily available in the literature. In doing so, we also set the stage for easily computing higher order relativistic corrections to these Newtonian effects.

Perhaps most importantly, we have described in detail how the RQF approach provides a deeper, and entirely new way of understanding tidal interactions, not in terms of a gravitational force, but in terms of a new, very simple, universal mechanism that is an exact, quasilocal manifestation of the equivalence principle in general relativity. In earlier papers~\cite{EMM2012,EMM2013} we described this mechanism in the context of transfer of energy and linear momentum through a system boundary; the extension to angular momentum is new in this paper. The mechanism is based on the simple fact that a mass gains energy and momentum when you accelerate towards it. More generally, the energy, linear momentum or angular momentum of a system can be changed by subjecting the frame in which we are viewing the system to linear or angular accelerations. In special relativity these non-inertial effects are accounted for by bulk terms in the conservation laws. In general relativity, these bulk terms become surface fluxes of gravitational energy, linear momentum and angular momentum passing through the system boundary. As emphasized in earlier papers, this shows that general relativistic effects such as frame-dragging are not just ``negligible corrections" to everyday physics, but are essential to explaining, at a deeper level, what is actually happening when, for example, we drop an apple. In the present paper we saw how tidal interactions (transfer of angular momentum through the system boundary) can be understood entirely in terms of the angular acceleration of RQF observers relative to the quasilocal mass-energy measured at the system boundary. This angular acceleration is what is required to keep the quasilocal frame {\it rigid} as gravitational angular momentum passes through the boundary, and is at the heart of the RQF approach. The RQF approach says that this mechanism is exact, applying even in the strong field regime; having shown in this paper that this mechanism reproduces well-known results in the context of tidal interactions in the 1PN limit provides further evidence of the viability of the RQF approach in both practical and theoretical applications.

\section*{Acknowledgements}

This work was supported in part by the Natural Sciences and Engineering Research Council of Canada.  We would also like to thank Eric Poisson for clearing up confusion surrounding the definition of the Love number.

\end{document}